\documentclass[aps,amsmath,amssymb,letterpaper,twocolumn,nofootinbib,superscriptaddress]{revtex4}
\pdfoutput=1

\usepackage{etoolbox}
\makeatletter
\patchcmd{\@ssect@ltx}
    {\addcontentsline{toc}{#1}{\protect\numberline{}#8}}
    {}
    {}
    {}
\makeatother
\usepackage[toc]{appendix}
\usepackage[hidelinks]{hyperref}

\makeatletter
\renewcommand{\p@subsection}{\thesection.}
\makeatother
\usepackage{graphicx}
\usepackage{epstopdf}
\usepackage{adjustbox}
\newcommand{\aligntop}[1]{\adjustbox{valign=t}{#1}}

\usepackage{paralist}
\newenvironment{myenuma}{\begin{inparaenum}[A.]}{\end{inparaenum}}
\newcommand{\dr}{\mathrm{d}}
\newcommand{\ir}{\mathrm{i}}
\newcommand{\e}{\mathrm{e}}
\DeclareMathOperator{\tr}{tr}

\DeclareMathOperator{\var}{Var}
\DeclareMathOperator{\sech}{sech}
\renewcommand{\Re}{\operatorname{\mathbb{R}e}}
\newcommand{\cdt}{\!\cdot}
\newcommand{\prn}[1]{\left ( #1 \right )}

\newcommand{\brk}[1]{\left [ #1 \right ]}
\newcommand{\abs}[1]{\left\lvert #1 \right\rvert}
\newcommand{\nrm}[1]{\left\lVert #1 \right\rVert}
\newcommand{\av}[1]{\left\langle #1 \right\rangle}
\newcommand{\set}[2]{\left\{ #1 \,\middle |\, #2 \right\}}
\newcommand{\diff}[3][\rule{0mm}{0mm}]{\frac{\mathrm{d}^{#1} #2}{\mathrm{d}{#3}^{#1}}}
\newcommand{\pdiff}[2]{\frac{\partial #1}{\partial #2}}

\newcommand{\intd}[2][\rule{0mm}{0mm}]{\int #1\!\!\dr #2\,}
\newcommand{\ie}{i.e.\ }

\newcommand{\etc}{etc.\ }
\newcommand{\wrt}{wrt.\ }
\newcommand{\CE}{\mathcal{E}}
\newcommand{\CO}{\mathcal{O}}
\newcommand{\ld}{\dot{\lambda}}
\newcommand{\kt}{k_\mathrm{B}T}
\newcommand{\thermmet}{g}
\newcommand{\fisher}{F}

\newcommand{\precest}{\operatorname{Prec}(\hat{\lambda})}
\newcommand{\preccov}{\Phi}
\newcommand{\spd}{V}
\newcommand{\mspd}{\overline{\spd}}
\newcommand{\pow}{\mathcal{P}_\text{ex}}
\newcommand{\covest}{\Sigma}
\newcommand{\varest}{\var(\hat{\lambda})}
\newcommand{\tmin}{\tau_\text{min}}
\newcommand{\tmax}{\tau_\text{max}}

\newcommand{\nump}{n}
\newcommand{\on}{^\text{on}}
\newcommand{\off}{^\text{off}}
\newcommand{\inv}{^{-1}}
\newcommand{\trans}{^\mathrm{T}}
\newcommand{\I}{\mathbf{I}}
\newcommand{\dinv}{^\mathrm{D}}
\newcommand{\onev}{\mathbf{e}}
\newcommand{\zerov}{\boldsymbol{0}}
\newcommand{\MM}{K}
\newcommand{\pr}{\mathbf{p}}
\newcommand{\eq}{\pi}
\newcommand{\Eq}{\Pi}
\newcommand{\fund}{Z}
\newcommand{\arow}{\boldsymbol{\xi}}

\newcommand{\uv}{\mathbf{u}}
\newcommand{\vv}{\mathbf{v}}
\newcommand{\evr}{u}
\newcommand{\evl}{\eta}

\newcommand{\tauh}{\widetilde{\tau}}
\newcommand{\tminh}{\tauh_\text{min}}
\newcommand{\fundh}{\widetilde{\fund}}

\newcommand{\spdb}{\mathbf{\spd}}
\newcommand{\mspdb}{\overline{\spdb}}

\newcommand{\fisherb}{\mathbf{\fisher}}

\newcommand{\covestb}{\boldsymbol{\covest}}
\newcommand{\preccovb}{\boldsymbol{\preccov}}
\newcommand{\eqb}{\boldsymbol{\eq}}
\newcommand{\Eqb}{\boldsymbol{\Eq}}
\newcommand{\evrb}{\mathbf{\evr}}
\newcommand{\evlb}{\boldsymbol{\evl}}
\newcommand{\phib}{\boldsymbol{\phi}}
\newcommand{\MMb}{\mathbf{\MM}}
\newcommand{\fundb}{\mathbf{\fund}}
\newcommand{\fundhb}{\widetilde{\fundb}}
\newcommand{\evrhb}{\widetilde{\evrb}}
\newcommand{\evlhb}{\widetilde{\evlb}}
\newcommand{\MMhb}{\widetilde{\MMb}}
\newcommand{\bl}{\boldsymbol{\lambda}}
\newcommand{\tlambda}{\tilde{\lambda}}
\newcommand{\tld}{\dot{\tlambda}}
\newcommand{\tlh}{\hat{\tlambda}}
\newcommand{\varesth}{\var\!\prn{\tlh}}
\newcommand{\precesth}{\operatorname{Prec}\!\prn{\tlh}}
\newcommand{\covesth}{\widetilde{\covest}}
\newcommand{\preccovh}{\widetilde{\preccov}}
\newcommand{\fisherh}{\widetilde{\fisher}}

\begin{document}
%
\title{A universal tradeoff between power, precision and speed in physical communication}
\author{Subhaneil Lahiri}
\affiliation{Department of Applied Physics, Stanford University, Stanford CA 94305, USA}
\author{Jascha Sohl-Dickstein}
\affiliation{Department of Applied Physics, Stanford University, Stanford CA 94305, USA}
\affiliation{Google Research, Mountain View CA 94043, USA}
\author{Surya Ganguli}
\affiliation{Department of Applied Physics, Stanford University, Stanford CA 94305, USA}
\begin{abstract}
  Maximizing the speed and precision of communication while minimizing power dissipation is a fundamental engineering design goal.
  Also, biological systems achieve remarkable speed, precision and power efficiency using poorly understood physical design principles.
  Powerful theories like information theory and thermodynamics do not provide general limits on power, precision and speed.
  Here we go beyond these classical theories to prove that the product of precision and speed is universally bounded by power dissipation in any physical communication channel whose dynamics is faster than that of the signal.
  Moreover, our derivation involves a novel connection between friction and information geometry.
  These results may yield insight into both the engineering design of communication devices and the structure and function of biological signaling systems.
\end{abstract}
\maketitle

\tableofcontents

\section{Introduction}\label{sec:intro}

Evolution has discovered remarkably rapid, precise, and energy efficient mechanisms for biological computation.
For example the human brain performs myriad computations, including complex object recognition in 150 ms \cite{Thorpe1996speed} while consuming less than 20 watts \cite{sterling2015principles}.
In contrast supercomputers operate in the megawatt range, and cannot yet rival general human performance.
To both understand the design principles governing biological computation, and to exploit such principles in engineered systems, it is essential to develop a general theoretical understanding of the relationship between power, precision and speed in computation.
For example, what are the fundamental limits and design tradeoffs involved in simultaneously optimizing these three quantities?

\par
Existing general theories, while powerful, often elucidate fundamental limits on at most two of these quantities.
For example, information theory \cite{Shannon1948infotheory,cover2012infotheory} provides limits on the accuracy of communication under power constraints;
but achieving these limits may require coding messages in asymptotically large blocks, thus providing no theoretical guarantees on speed (though see recent work \cite{Gastpar2003code,Polyanskiy2010dispersion} on finite block length coding).
Thermodynamics, through the second law, places fundamental limits on the work required to implement a physical process;
however, achieving such limits on thermodynamic efficiency requires quasistatic processes that unfold infinitely slowly.
More recent work has elucidated the minimal energy required to perform a physical process in finite time \cite{Schmiedl2007finitetime,Then2008finitetime}), but does not address accuracy in any computation.
Landauer \cite{Landauer1961irrev,Bennett1973irrev,Bennett1982review} revealed that the erasure of information sets a lower bound on the energy consumed in computation.
This observation inspired reversible computing \cite{Toffoli1980reversible}, which can achieve accurate computation at asymptotically zero energy expenditure, but at the expense of requiring asymptotically infinite time in the presence of noise.

\par
In the absence of general theories governing performance limits of computation at finite power, precision and speed, many works in systems biology have focused on tradeoffs between subsets of these quantities in very specific chemical kinetic schemes for specific computations.  Fundamental work on kinetic proofreading studied two way trade-offs between energy and accuracy \cite{Hopfield1974,Savageau1979,Savageau1979a,Freter1980,Ehrenberg1980,Blomberg1981,Savageau1981,Qian2006,Murugan2014}  or speed and accuracy \cite{Murugan2012} in the communication of genetic information. Also, many works have studied specific tradeoffs between energy and precision in cellular chemosensation \cite{Endres2009,Mehta2012,Lang2014,Barato2014,Govern2014diss,Govern2014resource,Sartori2014}.
Notably, \cite{Lan2012} studied simultaneous tradeoffs between power, speed and accuracy, but again in a very specific scheme for sensory adaptation.

\par Here we derive a general three-way performance limit on power, precision and speed in physical communication.
We focus on the problem of communication as it is a fundamental prerequisite for more complex computations.
Indeed, in modern parallel computing, communication between processors is an essential bottleneck for energy efficiency \cite{Dally2004}.
Our derived performance limit applies to ${\it any}$ Markovian communication channel whose internal dynamics is faster than dynamics of the external signal to be communicated.
In such a scenario, the external signal drives the communication channel into a non-equilibrium regime, in which the power dissipated can be described through a thermodynamic friction tensor on a manifold of channel state distributions \cite{Sivak2012metric,Zulkowski2012geo,Zulkowski2013noneq,Zulkowski2014bit,Mandal2015nessmet}.  We derive a lower bound on this friction tensor in terms of Fisher information, a fundamental quantity in the geometry of information \cite{amari2007methods}.
By developing a novel inequality relating friction, which governs energy dissipation, to information geometry, which governs accuracy in statistical estimation, we derive our general relation between power, precision and speed.  In essence, we find that the product of precision and speed is bounded by power.


\section{Physical channels coupled to external signals}\label{sec:channel}
\begin{figure}[tbh]
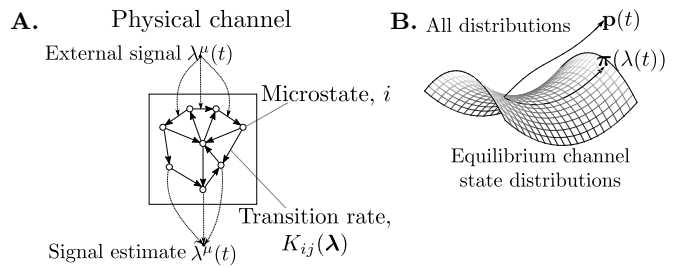

  \begin{flushleft}
  \begin{myenuma}
    {\bf\item\label{fig:channel}}\aligntop{\includegraphics[width=0.53\linewidth]{channel2.eps}}
    {\bf\item\label{fig:surf}}\aligntop{\includegraphics[width=0.37\linewidth]{surf.eps}}
  \end{myenuma}
  \end{flushleft}
  \caption{Modeling physical channels coupled to external signals.
  {\bf\ref{fig:channel}.}\ An external signal $\lambda^\mu(t)$ modulates the transition rates of an arbitrary continuous-time Markovian dynamical system, modeling a physical channel in contact with a heat bath.
  A downstream receiver can construct an estimate $\hat{\lambda}^\mu$ of the instantaneous signal by observing the instantaneous microstate of the channel.
  {\bf\ref{fig:surf}.}\ A manifold of equilibrium channel state distributions $\eqb(\bl)$ with intrinsic coordinates given by constant signal values $\bl$.
  Temporally varying signals $\bl(t)$ drive the  channel microstates through a trajectory of
  non-equilibrium distributions $\pr(t)$, off the equilibrium manifold.}
\label{fig:model}
\end{figure}

We model the communication channel as a physical system in contact with a thermal bath at inverse temperature $\beta=1/\kt$.
The channel is coupled to an $n$ dimensional signal $\bl$, specified by components $\lambda^\mu$, $\mu=1\ldots n$, so that the energy of the channel in microstate $i$ is $E_i(\bl)$.
When the external signal is held at a fixed $\bl$, we assume the channel relaxes to an equilibrium Boltzmann distribution
\begin{equation}\label{eq:canonical}
  \pi_i(\bl) = \e^{-\beta \left[ E_i(\bl) - \mathcal{F}(\bl) \right]},
\end{equation}
where $\mathcal{F}$ is the free energy.
We describe the non-equilibrium dynamics of the channel by a continuous-time Markov process, where the transition rate from state $i$ to state $j$ is $\MM_{ij}$ and $\MM_{ii} = - \sum_{j \neq i} \MM_{ij}$.
We assume the dynamics satisfies detailed balance:
\begin{equation}\label{eq:detbal}
  \eq_i(\lambda) \MM_{ij}(\lambda) = \eq_j(\lambda) \MM_{ji}(\lambda).
\end{equation}
Thus the external signal modifies the channel dynamics by directly modulating the transition rates (\autoref{fig:model}\ref{fig:channel}).

Under the dynamics in \eqref{eq:detbal}, for fixed external signal $\bl$, the channel state distribution relaxes to \eqref{eq:canonical}, yielding a manifold of equilibrium channel state distributions parameterized by $\bl$.
However, signals varying in time at a finite speed will drive the channel state distribution off the equilibrium manifold into a non-equilibrium distribution $\pr(t)$.
This distribution will be distinct from the equilibrium distribution $\eqb(\lambda(t))$ associated with the  instantaneous value of the external signal (\autoref{fig:model}\ref{fig:surf}).
By driving the channel at finite speed, temporally varying signals perform physical work on the channel.  Some of this work contributes to a change in free energy of the channel, while the rest is irreversibly dissipated as heat into the thermal bath.
Thus temporally varying signals yield a dissipation of excess power.

The non-equilibrium distribution $\pr(t)$ also contains information about the history of the signal $\bl(t)$.  Thus a downstream observer that can measure the channel microstate could use this information to estimate the signal with some level of precision, subject to a bound on signal speed.
Below we discuss in further detail the nature of signal speed, channel power dissipation, channel information geometry, and estimation precision, and we derive general relations between these quantities.
Moreover, in \ref{sec:ness} we discuss an extension of our results to situations where the channel dynamics breaks detailed balance, and the manifold of equilibrium distributions (\eqref{eq:canonical} and \autoref{fig:model}\ref{fig:surf}) is replaced with a manifold of non-equilibrium steady states.

\section{Power dissipation and the friction tensor}\label{sec:thermfisher}
In general, because the non-equilibrium distribution $\pr(t)$ depends on the entire history of the past temporal signal $\bl(t')$ for $t' < t$, the power dissipation due to a changing signal can also depend on this entire history.
However, if the temporal signal $\bl(t)$ varies more slowly than the channel dynamics (see \ref{sec:valid} for a precise description of this slow signal regime), then the non-equilibrium channel distribution $\pr(t)$ remains close
to the equilibrium manifold in \autoref{fig:model}\ref{fig:surf},  and the excess power dissipation at time $t$ depends on the signal history only through its instantaneous value $\bl(t)$ and time derivative $\dot{\bl}(t)$ \cite{Sivak2012metric}:
\begin{equation}\label{eq:diss}
  \pow =  \sum_{\mu\nu} \, \thermmet_{\mu\nu}(\bl) \, \ld^\mu \ld^\nu,
\end{equation}
where $\thermmet_{\mu\nu}$ is a friction tensor on the signal manifold,
\begin{equation}\label{eq:metricdef}
\begin{aligned}
  \thermmet_{\mu\nu}(\bl) &= \kt \intd[_0^\infty]{t'} \av{\delta\phi_\mu(0)\, \delta\phi_\nu(t')},\\
  \phi_\mu^i & = -\beta \pdiff{E_i}{\lambda^\mu},\qquad
  \delta\phi_\mu^i = \phi_\mu^i - \av{\phi_\mu}.
\end{aligned}
\end{equation}
Here expectations are computed with respect to the equilibrium distribution $\eqb(\bl)$, and derivatives are computed at the point $\bl$.
$\phi_\mu^i$ is the conjugate force exerted by the channel in response to changing a single signal component $\lambda^\mu$ when the channel is in microstate $i$.
Thus the statistics of force fluctuations at equilibrium combined with finite signal velocity determines excess power dissipation out of equilibrium, in the slow signaling limit.


\section{From friction to information geometry}\label{sec:fricgeo}
We now derive a lower bound on the friction tensor for the models of physical channels described above (see \ref{sec:friceigen} for more details).
First, the force correlation in \eqref{eq:metricdef} can be written as
\begin{equation}\label{eq:detbalcorr1}
  \av{\delta\phi_\mu(0)\delta\phi_\nu(t')} = \sum_{ij} p_{ij}(0,t') \delta\phi_\mu^i\delta\phi_\nu^j,
\end{equation}
where  $p_{ij}(t,t') = \eq_i \brk{ \exp (\MMb(\bl) \, (t'-t)) }_{ij}$ is the probability of being in state $i$ at time $t$ and in state $j$ at a later time $t'$, under equilibrium dynamics at a constant external signal $\bl$.

To simplify the matrix exponential, it is useful to employ an eigendecomposition of the rate matrix: $\MMb = -\sum_a q_a \evrb^a \evlb^a$.
Here $\evrb^a$ are column vectors obeying  $\MMb\evrb^a = -q_a \evrb^a$, $\evlb^a$ are row vectors obeying  $\evlb^a\MMb = -q_a \evlb^a$, and they further obey the normalization condition $\evlb^a \evrb^b = \delta^{ab}$.
With detailed balance, the eigenrates $q_a$ are real and positive, ordered in increasing order,
and the eigenvectors can be chosen to be real, satisfying $\evl^a_i = \eq_i \evr^a_i$.
The slowest eigenmode is the $a=0$ stationary mode, with $q_0=0$, $\evlb^0=\eqb$ and $\evrb^0=\onev$, a column vector of ones.
We assume that the Markov dynamics is ergodic, so the 0th eigenvalue of $\MMb$ is non-degenerate.

Now inserting \eqref{eq:detbalcorr1} into \eqref{eq:metricdef}, transforming to the eigenbasis of $\MMb$, and integrating over time $t'$ yields (see \ref{sec:friceigen})
\begin{equation}\label{eq:detbalmet}
\begin{aligned}
  \thermmet_{\mu\nu} &= \kt \sum_{a>0} \tau_a \prn{\evlb^a \cdt \delta\phib_\mu} \prn{\evlb^a \cdt \delta\phib_\nu} \\
     &\geq \kt \, \tmin \sum_{a>0} \prn{\evlb^a \cdt \delta\phib_\mu} \prn{\evlb^a \cdt \delta\phib_\nu}
     \\ &
     = \kt\, \tmin \fisher_{\mu\nu},
\end{aligned}
\end{equation}
where $\tau_a=1/q_a$, $\tmin = \min_{a>0} \tau_a$, and
\begin{equation}\label{eq:fisherdef}
  \fisher_{\mu\nu} =  \sum_i  \pi_i(\bl) \brk{ \partial_{\lambda_\mu } \ln \pi_i(\bl) } \brk{ \partial_{\lambda_\nu} \ln \pi_i(\bl) }
\end{equation}is the Fisher information.
The inequality \eqref{eq:detbalmet} means that $\thermmet_{\mu\nu} - \kt\, \tmin\fisher_{\mu\nu}$ is a positive semi-definite matrix.

\par This bound depends on the fastest channel timescale $\tmin$, and is only tight when the channel has a single timescale.
Systems with many degrees of freedom can often have extremely fast timescales.
However, in practice, signals do not couple to arbitrarily fast time-scales.
In this case, $\tmin$ should thought of as the fastest channel time-scale $\tau_a$ that is appreciably driven by the signal (i.e. $\evlb^a \cdt \delta\phib_\mu$ non-negligible).
Indeed, we will see below two examples where this timescale is much slower than the channel's fastest timescale.


\section{A power--precision--speed inequality}\label{sec:ppsbound}
The previous section revealed a simple inequality relating friction to information.
Here we build on this inequality to derive a general relation between power, precision and signal speed.
In particular, the Fisher information in \eqref{eq:fisherdef} is a Riemannianian metric on the manifold of equilibrium channel state distributions, describing the information geometry \cite{amari2007methods} of this manifold.
This metric measures the sensitivity of the channel distribution $\eqb(\bl)$ to changes in the signal $\bl$.
Intuitively, the higher this sensitivity, the more precisely one can estimate the signal $\bl$ from an observation of the stochastic microstate $i$ of the channel.

\par This intuition is captured by the Cramer-Rao theorem.
For simplicity, we focus below on the case of a one dimensional signal $\lambda$.
We derive analogous results for multidimensional signals in \ref{sec:multi}.
Consider a single observation of the channel microstate $i$, drawn from the equilibrium channel distribution $\eqb(\lambda)$.
Further consider an unbiased signal estimator $\hat{\lambda}(i)$, \ie a function of the stochastic channel microstate $i$ whose mean over observations is equal to the true signal $\lambda$.
The precision of this estimator is defined as the reciprocal of the variance of $\hat \lambda$ over the channel stochasticity:  $\precest = \frac{1}{\varest}$.
The Cramer-Rao \cite{Cramer1945,RadhakrishnaRao1945} bound states that estimator precision is bounded by Fisher information,
\begin{equation}\label{eq:cramerrao1}
  \precest \leq \fisher,
\end{equation}
for {\it any} unbiased estimator $\hat \lambda$ (here we have dropped the indices in $\fisher_{\mu\nu}$ for the special case of scalar signals).

\par A potential complication in the application of the classical Cramer-Rao bound for static signals $\lambda$, to our case of time varying signals $\lambda(t)$ is that the channel microstate $i$ is drawn from a non-equilibrium distribution $\pr(t)$, not the equilibrium distribution $\eqb(\lambda(t))$.
However, in the slow signal limit, which is related to an expansion in the temporal derivatives of $\lambda(t)$,
we can neglect the discrepancy between these two distributions,
as any such discrepancy only corrects higher order terms in this expansion  (see \eqref{eq:fisher_corr}).
Thus to the leading order in the slow signal expansion, in which the relation \eqref{eq:diss} between the friction tensor and power dissipation holds, we can replace the Fisher information of $\pr(t)$ with that of $\eqb(\lambda(t))$.

\par Now, with the careful analysis of the validity of the slow signal limit in hand, by simply combining the relation \eqref{eq:diss} between power, friction, and signal speed, the inequality \eqref{eq:detbalmet} relating friction to information,  and the inequality \eqref{eq:cramerrao1}, relating Fisher information to precision, we derive our central result relating power, precision and speed:
\begin{equation}\label{eq:scalarbound}
  \precest \, \spd \leq \frac{\pow}{\kt\,\tmin},
\end{equation}
where $V = \ld^2$ is the squared signal velocity.
Thus the product of two desirable quantities, the communication  precision and signal speed, is upper bounded by an undesirable quantity, the excess power dissipation.
This inequality uncovers the fundamental result that any attempt to communicate faster signals at fixed precision, or with higher precision at fixed signal speed, necessarily requires greater power dissipation.
Moreover, this relation applies universally to \emph{arbitrary} physical channels.

\par Saturating \eqref{eq:scalarbound} requires finding a statistically efficient unbiased estimator $\hat \lambda$ that saturates the Cram\'er-Rao bound \eqref{eq:cramerrao1}.
For the exponential family distributions that occur in statistical mechanics, we show how to construct such estimators for coordinates ``dual''  to $\bl$ (see \cite{amari2007methods}, \eqref{eq:dual} and \eqref{eq:dualest}).


\section{Example systems}\label{sec:examples}
We now illustrate the general relations derived above in specific examples.
Here we only summarize the results. Further details can be found in \ref{sec:examples_details}, as well as more examples involving \hyperref[sec:ising_details]{multidimensional signals} and \hyperref[sec:receptor]{channels violating detailed balance}.


\subsection{Heavily over-damped harmonic oscillator}\label{sec:damposc}
%
Consider a heavily damped particle in a viscous medium moving in a quadratic potential, where the external signal $\lambda(t)$ controls the position of the potential's minimum.
The particle position $x$ obeys a Langevin equation,
\begin{equation}\label{eq:damposc}
  \zeta \dot{x} = -\kappa(x - \lambda(t)) + \sqrt{2\zeta\kt} \xi(t),
\end{equation}
where $\zeta$ is the drag coefficient, $\kappa$ is the potential's spring constant and $\xi(t)$ is zero mean Gaussian white noise with $\av{\xi(t)\xi(t')} = \delta(t-t')$, reflecting fluctuations due to a thermal bath.
The distribution of $x(t)$ given any signal history $\lambda(t)$ is Gaussian with moments
\begin{equation}\label{eq:damposcmean}
\begin{aligned}
  & \av{x(t)} = \int_0^\infty \!\!\frac{\dr t'}{\tau} \, \e^{-t'/\tau} \lambda(t-t')
    = \sum_{n=0}^{\infty} \brk{-\tau\diff{}{t}}^n \lambda(t),
  \\
  & \av{\delta x(t) \delta x(t')} = \sigma^2 \e^{-\abs{t-t'}/\tau},
\end{aligned}
\end{equation}
where $\tau = \frac{\zeta}{\kappa}$ is the channel's relaxation timescale and $\sigma^2  = {\frac{\kt}{\kappa}}$ is the variance of the channel's equilibrium position fluctuations.
In the slow signal limit, where $\lambda(t)$ varies over timescales larger than $\tau$, the channel's mean position approximately tracks the signal:  $\av{x(t)} \approx \lambda(t)$.
More precisely, \eqref{eq:damposcmean} reveals that this slow signal limit is equivalent to neglecting higher order terms in a temporal derivative expansion.
This truncation is a good approximation when the temporal signal has negligible power at frequencies larger than $\frac{1}{\tau}$ (see \ref{sec:valid}).

In this limit, a good estimator for the signal based on the channel state $x$ is simply $\hat{\lambda} = x$, and its precision is $\precest = \frac{1}{\sigma^2}$. In the same slow limit, we compute power dissipation (see \ref{sec:damposc_details}):
\begin{equation}\label{eq:damposcpow}
\begin{aligned}
  \pow &= \kappa \ld(t) \int_0^\infty \!\!\dr t' \, \e^{-t'/\tau} \ld(t-t')
   \approx \zeta \ld(t)^2.
\end{aligned}
\end{equation}
Intuitively, the drag force is given by $-\zeta\dot{x}$, so the rate of doing work against it is $\zeta\dot{x}^2$, and in the slow signal limit,
$x(t) \approx \lambda(t)$.
Finally, using the Fokker-Planck description of the channel (see \ref{sec:damposc_details}), we find the intrinsic channel eigenmode timescales are $\tau_n = \tau/n$, for $n=1,2,\ldots\infty$.
However, $\lambda$ only couples to the $n=1$ mode, so $\tmin=\tau$.

Combining all these results yields,
\begin{equation}\label{eq:damposcbnd}
  \frac{\precest \, \spd}{\pow} = \frac{[\sigma^{-2}] [\ld^2]}{[\zeta \ld^2]} = \frac{1}{\kt \, \tmin},
\end{equation}
revealing that the damped harmonic oscillator channel saturates the general bound \eqref{eq:scalarbound}.
Note that the precision, and the Fisher information, are given by $\frac{1}{\sigma^2} = \frac{\kappa}{\kt}$.
This means that increasing the spring constant, $\kappa$, increases precision, as it forces $x$ to track $\lambda$ more closely.
However, it also speeds up the system, \ie it decreases $\tau = \frac{\zeta}{\kappa}$, and has no net effect on the power consumption, $\pow = \zeta \ld^2$.
In contrast, increasing the drag coefficient, $\zeta$, will increase power consumption and slow down the system, as expected, but has no effect on precision.
In practice, it is not possible to make $\kappa$ arbitrarily large, or $\zeta$ arbitrarily small.
This will limit how small one could make $\tau$.


\subsection{Ising ring}\label{sec:ising}
Consider a one dimensional Ising ring with periodic boundary conditions, \ie $N$ spins, $\sigma_n = \pm 1$, with $\sigma_0=\sigma_N$, all receiving a signal $h$, with Hamiltonian
\begin{equation}\label{eq:isingham}
  H = - h \sum_n \sigma_n - J \sum_n \sigma_n \sigma_{n+1}.
\end{equation}
We perform all computations as the signal $h$ passes through $h=0$ at finite velocity $\dot h$, and we assume Glauber dynamics \cite{Glauber1963} (see \ref{sec:ising_details}) for the spins.
This channel can model cooperativity between cell surface chemical receptors \cite{Skoge2011coop}, with $\sigma_n = \pm 1$ representing the active and inactive receptor states, the field $h$ determined by the ligand concentration, and $J$ controlling receptor cooperativity.
Equivalently, this channel could model cooperatively in the opening and closing of voltage gated ion channels, with $h$ reflecting time-varying voltage and the spins reflecting channel configurations.

First, although the Ising ring Glauber dynamics has a spectrum of eigenmode timescales, with the shortest being $\frac{1}{\alpha N}$, where $\alpha$ is the overall rate of the dynamics, the signal $h$ couples only to a mode with a single timescale (see \ref{sec:ising_details}), yielding
\begin{equation}\label{eq:ringtmin}
  \tmin = \frac{\e^{2\beta J} \cosh 2\beta J}{\alpha}.
\end{equation}
This quantity increases with $J$, due to critical slowing down \cite{Hohenberg1977critdyn}.
The slow signal limit is valid when the timescale $\tau_h$ over which $h$ varies is much larger than $\tmin$.
For a fixed $\tau_h$, this limit yields an upper limit on values of $J$ that we can analyze, which is roughly $ J \ll \kt \ln \alpha \, \tau_h$.

The Fisher information is given by
\begin{equation}\label{eq:ringfisher}
  \fisher = N \beta^2\e^{2\beta J},
\end{equation}
which also increases with $J$.
In essence, increasing $J$ has two opposing effects on how well the spin statistics transmits the signal $h$.
First the gain of the mean spin response to $h$ (\ie magnetic susceptibility) increases, improving coding.
Second, the variance of spin response also increases with $J$, impairing coding.
The former effect dominates over the latter, leading to increased information with cooperativity.
Moreover, in \ref{sec:ising_details} we show how to construct an efficient unbiased estimator for the dual coordinate to the signal $h$ (see \eqref{eq:dual} and \eqref{eq:dualest}).

The power dissipated when $h$ is varied is given by
\begin{equation}\label{eq:ringpower}
   \pow = \frac{N \beta\, \e^{4\beta J} \cosh 2\beta J}{\alpha} \, \dot{h}^2.
\end{equation}
This also increases with $J$, partly due to increased response gain to $h$, and partly due to critical slowing down.
Now, combining \eqref{eq:ringtmin}, \eqref{eq:ringfisher}, and \eqref{eq:ringpower}, we find
\begin{equation}
  \frac{\fisher \, \spd}{\pow} = \frac{1}{(\kt) \, \tmin},
\end{equation}
where $V = \dot h^2$ is the squared signal velocity.
This implies the general bound \eqref{eq:scalarbound} would be saturated for the Ising ring if the Cramer-Rao bound \eqref{eq:cramerrao1} could be saturated.

\par
We note that while increasing $J$ increases Fisher information, the dissipated power increases even faster, yielding diminishing returns in terms of Fisher information per watt.
This is analogous to the diminishing returns exemplified by the concavity of the capacity-cost curve in information theory \cite{cover2012infotheory}.
Also, in \cite{Skoge2011coop}, the same system was analyzed in a different setting.
There the signal was static while the channel was observed for an extended period, whereas here the signal is changing and the channel is observed instantaneously.
There, increasing receptor cooperativity $J$ reduced performance, since critical slowing yields fewer independent signal observations.
Similarly, here we see that cooperativity also slows down the system, decreasing the right-hand-side of \eqref{eq:scalarbound}, as verified in \ref{sec:ising_details}.
In so doing it tightens the bound between power, precision and speed.


\section{Discussion}\label{sec:discussion}

In summary, by deriving general relations between friction and information, we have shown that the product of signal speed and channel precision cannot exceed power dissipation for an extremely general class of physical communication channels.
Intuitively, this three-way tradeoff arises because any increase in speed at fixed precision requires the channel state distribution to change more rapidly, leading to increased power dissipation.
Similarly any increase in precision at fixed speed requires high signal sensitivity, or a larger signal dependent change in the channel equilibrium state distribution as measured by the Fisher information metric, which again leads to greater power dissipation.

Our newly discovered three-way tradeoff motivates new experiments to assess exactly how close biological systems come to simultaneously optimizing power, precision and speed.
Indeed any experiment that measures only two of these three quantities fundamentally cannot assess how close evolution pushes biology to the limits set by physics in general information processing tasks.

Moreover, our work opens the door to intriguing theoretical extensions.
Here, we focused on tradeoffs in estimating the current value of a slowly changing signal from an instantaneous observation of a physical channel.
Alternatively, we could consider estimating either the temporal history of a signal from the instantaneous channel state \cite{Ganguli2008,Ganguli2010}, or estimating a static signal given an extended time series of channel states.
The former would involve Fisher information metrics of channel states over signal trajectories, while the later would involve the Fisher information of probability distributions over channel state trajectories.
It would be interesting to explore universal three-way tradeoffs between power, precision and speed in these more general dynamical scenarios.
We hope that the essential ideas underlying our mathematical derivation of a universal tradeoff between power precision and speed will be of benefit in understanding even more general scenarios of communication and computation across both biology and engineering.


\section*{Acknowledgements}

We thank Madhu Advani, Gavin Crooks, Dibyendu Mandal and the participants of the Lineq reading group at UC Berkeley for useful discussions.
We thank Genentech, the Office of Naval Research, the Burroughs-Wellcome Fund, and the Alfred P.\ Sloan, James S.\ McDonnell, Simons, and McKnight Foundations for funding.

\section*{Appendices}
\begin{appendices}
\makeatletter
\renewcommand{\p@section}{Appendix }
\renewcommand{\p@subsection}{Appendix \thesection.}
\makeatother


\section{Summary}\label{sec:suppintro}

In this supplement, we provide the details underlying the results of the main text.
In \ref{sec:friceigen} we show a step-by-step derivation of \eqref{eq:detbalmet}.
Next, \ref{sec:valid} contains a precise description of the slow signaling limit where the friction tensor used in \eqref{eq:diss} determines power dissipation.
In \ref{sec:multi}, we generalize the power-precision-speed inequality of the \autoref{sec:ppsbound} to multidimensional signals.
In \ref{sec:ness}, we further extend these results to channels whose dynamics violate detailed balance (\eqref{eq:detbal}), and therefore have nonequilibrium steady states.
Then, \ref{sec:fisher_corr} provides the justification for replacing the Fisher information of the out of equilibrium state of the channel with the Fisher information of the equilibrium distribution in the Cram\'er-Rao bound \eqref{eq:cramerrao1}.
Following which, \ref{sec:dual} contains the definition of the dual coordinate system for the signal manifold, which proves to be necessary for finding estimators that can saturate the Cram\'er-Rao bound.
In \ref{sec:examples_details}, we provide the details behind the discussion of the example systems in \autoref{sec:examples}, including the \hyperref[sec:damposc_details]{heavily over-damped harmonic oscillator}, the \hyperref[sec:ising_details]{Ising ring}, which we also extend to the case of multidimensional signals, and a \hyperref[sec:receptor]{four state receptor} that violates detailed balance.


\section{Relating friction and information geometry}\label{sec:friceigen}

In this section we provide a detailed derivation of \eqref{eq:detbalmet}.
We introduce an eigenvector basis
\begin{equation}\label{eq:markoveig}
  \MMb\evrb^a = -q_a \evrb^a, \qquad
  \evlb^a\MMb = -q_a \evlb^a,
\end{equation}
with $q_0=0$, $\evlb^0=\eqb$ and $\evrb^0=\onev$, a column vector of ones.
If we have detailed balance, then all $q_a, \evrb^a, \evlb^a$ are real and with suitable choice of normalization we have:
\begin{equation}\label{eq:detbaleig}
\begin{gathered}
  \evlb^a \evrb^b = \delta^{ab}, \quad
  \MMb = -\sum_a q_a \evrb^a \evlb^a, \\
  \I = \sum_a \evrb^a \evlb^a, \quad
  \evl^a_i = \eq_i \evr^a_i,
\end{gathered}
\end{equation}
where the last relation holds element-by-element.

The probability of being in state $i$ at time $t$ and in state $j$ at a later time $t'$ is
\begin{equation*}
\begin{aligned}
  p_{ij}(t,t') 
                 &=  \eq_i \brk{ \exp (\MMb(t'-t)) }_{ij} \\
                 &= \eq_i \sum_a \evr^a_i \, \e^{-q_a (t-t')} \, \evl^a_j,
\end{aligned}
\end{equation*}
assuming that the system is in equilibrium.
Then the correlation function in \eqref{eq:metricdef} can be written as
\begin{equation*}
\begin{aligned}
  \av{\delta\phi_\mu(0)\delta\phi_\nu(t')}
   &= \sum_{ij} p_{ij}(0,t') \delta\phi_\mu^i\delta\phi_\nu^j \\
   &= \sum_{ij}  \eq_i \brk{ \exp (\MMb t') }_{ij} \delta\phi_\mu^i\delta\phi_\nu^j \\
   &= \sum_{ij}\sum_a \eq_i \evr^a_i \,\e^{-q_a t'} \evl^a_j \,  \delta\phi_\mu^i\delta\phi_\nu^j \\
   &= \sum_{ij}\sum_a \evl^a_i \,\e^{-q_a t'} \, \evl^a_j \delta\phi_\mu^i\delta\phi_\nu^j \\
   &= \sum_{ij}\sum_a \e^{-q_a t'} \prn{\evl^a_i \delta\phi_\mu^i} \prn{\evl^a_j \delta\phi_\nu^j}.
\end{aligned}
\end{equation*}
Substituting this into \eqref{eq:metricdef} and integrating with respect to $t'$ leads to
\begin{equation*}
  \thermmet_{\mu\nu}(\bl) = \kt \sum_a \frac{1}{q_a} \sum_{ij} \prn{\evl^a_i \delta\phi_\mu^i} \prn{\evl^a_j \delta\phi_\nu^j}.
\end{equation*}
Defining $\tau_a=\frac{1}{q_a}$, this is the first line of \eqref{eq:detbalmet}.

To derive the last line of \eqref{eq:detbalmet}, we reverse the process above:
\begin{equation*}
\begin{aligned}
  \sum_a \sum_{ij} \prn{\evl^a_i \delta\phi_\mu^i} \prn{\evl^a_j \delta\phi_\nu^j}
    &=  \sum_{ij}\sum_a \evl^a_i  \evl^a_j \, \delta\phi_\mu^i\delta\phi_\nu^j \\
    &= \sum_{ij}\sum_a \eq_i\,  \evr^a_i  \evl^a_j \, \delta\phi_\mu^i\delta\phi_\nu^j \\
    &= \sum_{ij} \eq_i \, \delta_{ij} \, \delta\phi_\mu^i \delta\phi_\nu^j \\
    &= \sum_{i} \eq_i \, \delta\phi_\mu^i \delta\phi_\nu^i \\
    &= \fisher_{\mu\nu},
\end{aligned}
\end{equation*}
where we used the fact that, for the canonical ensemble \eqref{eq:canonical}, we have $\pdiff{\ln\eq_i}{\lambda^\mu} = \delta\phi^i_\mu$, so the penultimate line above is identical to \eqref{eq:fisherdef}.


\section{Domain of validity of the slow signal approximation}\label{sec:valid}

In this section, we will discuss the regime where the approximations used in the derivation of \eqref{eq:diss} are valid.

In the original derivation, \cite{Sivak2012metric}, the linear response approximation was used, which assumes that all changes in the parameters, $\lambda^\mu$, are small.
It was noted in \cite{Zulkowski2012geo} that the same result arises from truncating a derivative expansion at first order.
This was made explicit by a systematic derivative expansion in \cite{Mandal2015nessmet}.

One can legitimately truncate the derivative expansion at leading order when the functions $\lambda^\mu(t)$ vary slowly relative to the intrinsic timescales of the system.
This can be made precise by rewriting all expressions in terms of the Fourier transforms, $\widetilde{\lambda}^\mu(\omega)$.

In the derivative expansion in \cite{Mandal2015nessmet}, every time derivative comes with a factor of $\MMb\dinv$, the Drazin pseudoinverse of $\MMb$,  which can be defined in terms of the eigenvalues and eigenvectors in \eqref{eq:markoveig}:
\begin{equation}\label{eq:drazin}
  \MMb = - \sum_a q_a \evrb^a \evlb^a
  \; \Longleftrightarrow \;
  \MMb\dinv = - \sum_{a>0} \tau_a  \evrb^a \evlb^a,
\end{equation}
where $\tau_a = 1/q_a$.
Therefore, the eigenvalues of $\MM\dinv$ are the $\tau_a$.

Using  $\Delta\pr(t) = \pr(t) - \eqb(t)$, we can write the Fourier transform of the derivative expansion in \cite{Mandal2015nessmet} as
\begin{multline}\label{eq:derexp}
  \Delta\pr(\omega_0)
   = \sum_{n=1}^\infty \int\!\!\frac{\dr\omega_1}{2\pi} \cdots \int\!\!\frac{\dr\omega_n}{2\pi}\,
     \eqb(\omega_n)\,\times \\
     (-\ir\omega_n) \MMb\dinv(\omega_{n-1} - \omega_n) \cdots (-\ir\omega_1) \MMb\dinv(\omega_0-\omega_1).
\end{multline}
As each power of $\omega$ comes with one power of $\MM\dinv$, whose eigenvalues are $\tau_a$, the derivative expansion can be thought of as an expansion in $\omega\tau_a$.

We define $\omega_\text{max}$ to be the largest $\omega$ for which $\lambda^\mu(\omega)$ is significantly nonzero.
As $\eqb(t)$ and $\MMb(t)$ inherit their time dependence from $\lambda^\mu(t)$, this will also be of the same order of magnitude as the largest frequency for which $\eqb(\omega)$ and $\MMb\dinv(\omega)$ are significantly nonzero.
Then all integrals over $\omega$ will be dominated by regions that satisfy $\abs{\omega} \lesssim \CO(\omega_\text{max})$.
We also define $\tmax = \max_{a>0} \tau_a$.
This means that, provided that
\begin{equation}\label{eq:valid}
  \omega_\text{max} \tmax \ll 1,
\end{equation}
all factors of $\omega\tau_a$ that contribute to $\Delta\pr$ will be small and the expansion can be truncated at leading order.

As shown in \cite{Mandal2015nessmet}, this expression for $\Delta\pr$ can be used to compute the excess power dissipation via
\begin{equation*}
  \pow = T\diff{S}{t} + \kt \sum_{ij} \Delta p_i(t) \MM_{ij}(t) \ln \eq_j(t).
\end{equation*}
Using \eqref{eq:derexp}, this can be expressed as an expansion in $\omega\tau_a$ as well.
When \eqref{eq:valid} is satisfied, this expansion can be truncated at leading order, which was shown in \cite{Mandal2015nessmet} to be the $n=2$ term.


\section{Power precision speed tradeoff for multidimensional signals}\label{sec:multi}

In this section, we will extend the results of \autoref{sec:ppsbound} to systems with more than one varying parameter.

Suppose we have several parameters, $\lambda^\mu$, with a set of unbiased estimators, $\hat{\lambda}^\mu$, for all of them.
The Cram\'er-Rao bound states that the covariance of these estimators, $\covestb$, is bounded from below by the inverse Fisher information.
This can be rephrased in terms of the precision, $\preccovb$, defined as
\begin{equation*}
   \covest^{\mu\nu} = \av{\hat{\lambda}^\mu\hat{\lambda}^\nu}-\av{\hat{\lambda}^\mu}\av{\hat{\lambda}^\nu},
   \qquad
   \preccovb = \covestb\inv.
\end{equation*}
Then the Cram\'er-Rao bound can be written as:
\begin{equation}\label{eq:cramerrao}
  \preccovb \leq \fisherb,
\end{equation}
The diagonal elements of $\covestb$ are the squared uncertainties.

We can combine this with \eqref{eq:diss} and \eqref{eq:detbalmet} to prove the bound
\begin{equation}\label{eq:matrixbound}
  \tr(\preccovb\spdb) \leq \frac{\pow}{(\kt) \, \tmin} .
\end{equation}
%
Much like \eqref{eq:scalarbound}, this inequality bounds measures of precision (inverse variance of parameter estimates), $\preccovb$, and speed (squared parameter velocity), $\spdb$, both desirable properties of a communication channel, in terms of a measure of power use, $\pow$, an undesirable property of a channel.
This bound also applies to an \emph{arbitrary} physical system.

We can rewrite this bound in a different form for average quantities.
In general, there will be an ensemble of signals to be sent, resulting in an ensemble of trajectories for the control parameters, $\lambda^\mu(t)$.
This means that, at any instant of time, there will be a probability distribution for the current values of $\lambda^\mu$ and $\ld^\mu$.
We can obtain a bound on the instantaneous power dissipation as a function of the current value of $\lambda^\mu$,
\begin{equation*}
  \av{\pow} \geq  (\kt) \, \tmin \tr (\covestb\inv \mspdb),
  \quad \text{where} \quad
  \mspdb = \av{\spdb},
\end{equation*}
where the averages are over $\ld^\mu$ conditioned on $\lambda^\mu$.
This leads to the inequality
\begin{equation}\label{eq:ppsbound}
\begin{aligned}
  \av{\pow} \tr (\covestb \mspdb\inv)
   &\geq (\kt) \, \tmin \tr (\covestb\inv \mspdb) \tr (\covestb \mspdb\inv)
   \\&\geq \nump^2 (\kt) \, \tmin,
\end{aligned}
\end{equation}
where $\nump$ is the number of parameters.
We used the fact that $\covestb$ and $\mspdb$ are positive-definite,
therefore $\covestb \mspdb\inv$ is a similarity transform of the positive definite quantity $\mspdb^{-1/2} \covestb \mspdb^{-1/2}$, and therefore has positive eigenvalues.
Such matrices satisfy the inequality $\tr(\mathbf{A}) \tr(\mathbf{A}\inv) \geq \nump^2$, which can be seen by writing the traces as sums of eigenvalues.
This is saturated when $\mspdb \propto \covestb$.

This inequality lower bounds the product of measures of power use, $\pow$, imprecision (variance of parameter estimates), $\covestb$, and slowness (inverse mean squared parameter velocity), $\mspdb\inv$, for an \emph{arbitrary} physical system.
These are all undesirable properties of a communication system.
If we wish to lower any one of these, at least one of the others will necessarily increase for a system that achieves this bound.


\section{Communication through channels with non-equilibrium steady states}\label{sec:ness}

Here we will show how to extend the bound in \eqref{eq:detbalmet} to systems that do not satisfy detailed balance.
First we will need to define some mathematical quantities.

The natural inner product on the state space of an ergodic Markov process is the $\mathcal{L}^2_{\eqb}$ inner product, using the steady state distribution, $\eqb$ to weight states:
\begin{equation}\label{eq:innerprod}
  (\uv,\vv) \equiv \sum_i \eq_i u_i^* v_i = (\uv^*)\trans \Eqb \vv,
\end{equation}
where $\Eqb$ is a diagonal matrix with $\Eq_{ii}=\eq_i$.

The usual notion of transposing (or hermitian conjugating) vectors and matrices is only appropriate on spaces with an Euclidean inner product.
Here we define a different adjoint.
For a column vector $\uv$, a row vector $\arow$ and a matrix $\mathbf{M}$, the adjoints must satisfy the following relations:
\begin{equation*}
\begin{gathered}
  \uv^\dag \vv = (\uv,\vv),
  \qquad
  (\arow^\dag,\uv) = \arow\uv,
  \\
  (\mathbf{M}^\dag\uv,\vv) = (\uv,\mathbf{M}\vv).
\end{gathered}
\end{equation*}
The adjoints are given by:
\begin{equation}\label{eq:adjoint}
\begin{gathered}
  \uv^\dag = (\uv^*)\trans \Eqb,
  \qquad
  \arow^\dag = \Eqb\inv (\arow^*)\trans ,
  \\
  \mathbf{M}^\dag = \Eqb\inv (\mathbf{M}^*)\trans \Eqb.
\end{gathered}
\end{equation}
When applied to transition matrices, we see that the adjoint is time-reversal.

In particular, this means that detailed balance is equivalent to the self-adjointness of $\MMb=\MMb^\dagger$, which implies that its eigenvalues are real and its eigenvectors can be chosen to be real and orthonormal under the inner product \eqref{eq:innerprod}.
Taking the adjoint of the eigenvector equation \eqref{eq:markoveig} then tells us that $\evlb^a = (\evrb^a)^\dag$ (with suitable choice of normalisation) when detailed balance is satisfied, which implies the last formula in \eqref{eq:detbaleig}.

The transition matrix $\MMb$ is not invertible, as $\eqb$ and $\onev$ are left/right null vectors.
Instead, we can use the fundamental matrix of the Markov process, defined as \cite{kemeny1960finite}%
\footnote{The fundamental matrix, $\fundb$, is related to the Drazin pseudoinverse in \eqref{eq:drazin} by $\MMb\dinv = \tau\onev\eqb - \fundb$.}
\begin{equation}\label{eq:funddef}
  \fundb = \prn{\frac{\onev\eqb}{\tau} - \MMb}\inv,
\end{equation}
where $\tau$ is an arbitrary finite positive scalar with units of time.
Nothing we calculate will depend on its value.
One can show that
\begin{equation*}
  \eqb\fundb=\tau\eqb, \qquad
  \fundb\onev=\tau\onev, \qquad
  \fundb\MMb = \onev\eqb - \I.
\end{equation*}
The fundamental matrix, thus defined, has the same eigenvectors as $\MM$
\begin{equation*}
  \fundb = \tau\onev\eqb + \sum_{a>0} \tau_a \evrb^a \evlb^a.
\end{equation*}

For systems that have no energy function, we can write the ``force'', $\phi$, in terms of the steady state distribution \cite{Hatano2001,Zulkowski2013noneq}
\begin{equation*}
  \delta\phi^i_\mu = \pdiff{\ln\eq_i}{\lambda^\mu} .
\end{equation*}
Its correlation function can be written as
\begin{equation*}
\begin{aligned}
  \av{\delta\phi_\mu(0)\delta\phi_\nu(t')}
    &= \sum_{ij} p_{ij}(0,t') \delta\phi_\mu^i\delta\phi_\nu^j
    \\&= -\diff{}{t'}\sum_{ij} \eq_i \delta\phi_\mu^i \brk{ \fund\e^{\MM t'}  }_{ij} \delta\phi_\nu^j.
\end{aligned}
\end{equation*}
Then the friction tensor is given by
\begin{equation*}
\begin{aligned}
  \thermmet_{\mu\nu} &= \kt \intd[_0^\infty]{t'} \av{\delta\phi_\mu(0)\delta\phi_\nu(t')}
   \\&= \kt \sum_{ij} \eq_i \fund_{ij} \delta\phi_\mu^i \delta\phi_\nu^j .
\end{aligned}
\end{equation*}
A similar result can be found in \cite{Mandal2015nessmet}.

Noting that we need only keep the symmetric part, as the antisymmetric part will not contribute to the dissipation, as it is traced with a symmetric tensor, $\ld^\mu\ld^\nu$, in \eqref{eq:diss}, we can replace this with
\begin{equation}\label{eq:metricgen}
  \thermmet_{\mu\nu} = \frac{\kt}{2}\sum_{ij} \prn{ \eq_i \fund_{ij} + \eq_j \fund_{ji} }\delta\phi_\mu^i \delta\phi_\nu^j.
\end{equation}
Armed with our definition of the adjoint from \eqref{eq:adjoint}, we can rewrite \eqref{eq:metricgen} as
\begin{equation*}
\begin{aligned}
  \thermmet_{\mu\nu} &= (\kt) \sum_{ij} \eq_i \prn{ \frac{\fund_{ij} + \fund_{ij}^\dag}{2} }\delta\phi_\mu^i \delta\phi_\nu^j
             \\&= \kt \sum_{ij} \eq_i \fundh_{ij} \delta\phi_\mu^i \delta\phi_\nu^j.
\end{aligned}
\end{equation*}
where we introduced a new matrix $\fundhb$%
\footnote{Note that $\fundhb$ is not necessarily the fundamental matrix of some Markov process.
While the putative transition rate matrix, $\MMhb = \frac{\onev\eqb}{\tau} - \fundhb\inv$, does indeed have zero row sums, we do not know if the off-diagonal elements are nonnegative in general.}
with the following eigenmode decomposition:

\begin{equation}\label{eq:fundsymev}
  \fundhb \equiv \frac{\fundb + \fundb^\dag}{2}
                  = \tau\onev\eqb + \sum_{a>0} \tauh_a \evrhb^a \evlhb^a.
\end{equation}
where $\tau,\tauh_a$ are the eigenvalues and $\onev,\evrhb^a$($\eqb,\evlhb^a$) are the right(left) eigenvectors of the self-adjoint matrix $\fundhb$.
This means that all these quantities are real, orthogonal and $\evlhb^a = (\evrhb^a)^\dag$.

Note that $\fundb(\MMb^\dag) = \fundb(\MMb)^\dag$.
Therefore, for systems that satisfy detailed balance, the symmetrization in \eqref{eq:metricgen} is unnecessary and $\fundhb=\fundb$.
For such systems, the eigenvectors and timescales that appear in \eqref{eq:fundsymev} are the eigenvectors and timescales of the original transition matrix.

Proceeding in analogy with \eqref{eq:detbalmet}, we can write
\begin{equation}\label{eq:metgeneig}
\begin{aligned}
  \thermmet_{\mu\nu} &= \kt \sum_{a>0} \tauh_a \prn{\evlhb^a \cdt \delta\phib_\mu} \prn{\evlhb^a \cdt \delta\phib_\nu}
     \\&\geq (\kt) \tminh \fisher_{\mu\nu}.
\end{aligned}
\end{equation}
The inequality is meant in the operator sense, \ie the difference between the left and right hand sides is positive semi-definite.
This is essentially the same as the case with detailed balance, only now the physical interpretation of $\tminh$ is less clear.

If we apply this to the power--precision--speed bounds \eqref{eq:scalarbound}, \eqref{eq:matrixbound} and \eqref{eq:ppsbound}, we find
\begin{equation}\label{eq:ppsboundgen}
\begin{aligned}
  \precest \, \spd &\leq \frac{\pow}{(\kt)\,\tminh},\\
    \tr(\preccovb\spdb) &\leq \frac{\pow}{(\kt) \, \tminh},\\
  \av{\pow} \tr (\covestb \mspdb\inv) &\geq \nump^2 (\kt) \, \tminh.
\end{aligned}
\end{equation}

Now we'll argue that $\tauh_a > 0$ without detailed balance.
This guarantees that the friction tensor is positive definite (the authors of \cite{Zulkowski2013noneq} were unsure if this was an issue).

The numerical range (a.k.a.\ field of values) of a matrix $\mathbf{M}$ over a subspace $\mathcal{V}$ is defined as
\begin{equation*}
  W(\mathbf{M},\mathcal{V}) = \set{ \frac{(\uv,\mathbf{M}\uv)}{\nrm{\uv}^2} }{ \uv\in\mathcal{V}, \uv\neq\zerov },
\end{equation*}
where the norm is computed with the inner product \eqref{eq:innerprod}.
Strictly speaking, we will be using at the closure of this set.
Setting $\mathcal{V}$ to the subspace orthogonal to $\onev$ under the inner product \eqref{eq:innerprod}:
\begin{equation*}
  \tauh_a \in W(\fundhb, \onev^\perp) = \Re W({\fundb}, \onev^\perp).
\end{equation*}
From the Cheeger inequality \cite{Lawler1988cheeger}, we know that the real part of the numerical range of $\MM$ is negative:%
\footnote{For non-normal $\MMb$ (\wrt the inner product \eqref{eq:innerprod}) this condition is stronger than the statement that its spectrum has negative real part. For normal $\MMb$, the closure of the numerical range is the convex hull of the spectrum and the two statements are equivalent.}
\begin{equation*}
  \Re W(\MMb, \onev^\perp) < 0.
\end{equation*}
As $\MM$ is invertible in the space $\onev^\perp$, we can write $\uv=\MMb\vv$, $\vv=-\fundb\uv$ in the following:
\begin{equation*}
\begin{gathered}
\begin{aligned}
  W({\fundhb}, \onev^\perp)
    &= \set{\Re \frac{(\uv,\fundb\uv)}{\nrm{\uv}^2} }{ \uv\in\onev^\perp, \uv\neq\zerov } \\
    &= \set{-\Re \frac{(\MMb\vv,\vv)}{\nrm{\MMb\vv}^2} }{ \vv\in\onev^\perp, \vv\neq\zerov } \\
    &= \set{-\Re \frac{(\vv,\MMb\vv)}{\nrm{\vv}^2}\frac{\nrm{\vv}^2}{\nrm{\MMb\vv}^2} }{ \vv\in\onev^\perp, \vv\neq\zerov } \\
    &> 0
\end{aligned}\\
    \qquad \text{as} \quad
    \frac{(\vv,\MMb\vv)}{\nrm{\vv}^2} \in W(\MMb, \onev^\perp)
    \quad \text{and} \quad
    \frac{\nrm{\vv}^2}{\nrm{\MMb\vv}^2} > 0.
\end{gathered}
\end{equation*}
This implies that $\tauh_a > 0$, and therefore the friction tensor \eqref{eq:metgeneig} is positive definite.


\section{Corrections to information-precision relations at finite signal speed}\label{sec:fisher_corr}

The inequality in \eqref{eq:detbalmet} contains the Fisher information of the equilibrium distribution for the current value of $\lambda^\mu(t)$.
However, when we used the Cram\'er-Rao bound above \eqref{eq:cramerrao}, we should use the actual probability distribution over the system's microstates.
This will lag behind the current equilibrium distribution when the parameters change at a nonzero speed.
As shown in \cite{Mandal2015nessmet}, the correction to the distribution can be expressed as a derivative expansion.
The leading term is
\begin{equation}\label{eq:prob_corr}
\begin{aligned}
  \pr(t) &= \eqb(t) + \Delta \pr(t),
  \\
  \Delta \pr(t) &= -\dot{\eqb}(t)\fundb(t) + \CO(\ddot{\lambda},\ld^2),
\end{aligned}
\end{equation}
where $\fundb$ is the fundamental matrix introduced in \eqref{eq:funddef}.

We can write the corrected Fisher information as
\begin{equation*}
  \fisher_{\mu\nu} = \sum_i p_i \pdiff{\ln p_i}{\lambda^\mu} \pdiff{\ln p_i}{\lambda^\nu}
  = \fisher^\text{eq}_{\mu\nu} + \Delta\fisher_{\mu\nu}.
\end{equation*}
Using \eqref{eq:prob_corr}, we find
\begin{equation}\label{eq:fisher_corr}
\begin{aligned}
  \Delta\fisher_{\mu\nu} =&\,
    -\ld^\rho \Biggl\{
      \sum_{ij} \eq_i \fund_{ij}
        \Biggl[(\delta\phi^i_{\mu\rho}+\delta\phi^i_\mu\delta\phi^i_\rho)\delta\phi^j_\nu
        \\&+(\delta\phi^i_{\nu\rho}+\delta\phi^i_\nu\delta\phi^i_\rho)\delta\phi^j_\mu - \delta\phi^i_\rho\delta\phi^j_\mu\delta\phi^j_\nu  \\
        &+\delta\phi^i_\rho \sum_{kl}  \fund_{kl}
          \prn{\partial_\mu \MM_{jk} \delta\phi^l_\nu + \partial_\nu \MM_{jk} \delta\phi^l_\mu} \Biggr]
          \Biggr\}
           \\&+ \CO(\ddot{\lambda},\ld^2),
\end{aligned}
\end{equation}
where $\phi^i_{\mu\nu} = \frac{\partial^2 E_i}{\partial\lambda^\mu \partial\lambda^\nu}$, $\delta\phi^i_{\mu\nu} = \phi^i_{\mu\nu} - \av{\phi_{\mu\nu}}$ and $\partial_\mu\MM_{ij} = \pdiff{\MM_{ij}}{\lambda^\mu}$.

We see that replacing the Fisher information of the equilibrium distribution in \eqref{eq:detbalmet} with the Fisher information of the current distribution results in corrections that are higher order in $\ld$ than the term that we have kept.


\section{Dual coordinates and optimal estimators}\label{sec:dual}

When the receiver of the signal attempts to reconstruct it, the aim is to find the correct point on the manifold of control parameters.
All coordinate systems on this manifold provide equally good descriptions of that point.
However, when it comes to constructing estimators of the coordinates themselves, some coordinate systems are better than others from the point of view of estimator bias and variance.

While the Cram\'er-Rao bound on the error in estimating the location of this point holds in \emph{any coordinate system}, it may be easier to construct unbiased estimators to saturate this bound in a specific coordinate system.
For the exponential family of equilibrium distributions, this privileged coordinate system is the dual coordinate system \cite{amari2007methods}.
If the energy is linearly dependent on the parameters $\lambda^\mu$, \ie $E = -\sum_\mu \lambda^\mu \CO_\mu$, the dual coordinates are defined by
\begin{equation}\label{eq:dual}
  \tlambda^\mu = \av{\CO_\mu} = -\av{\pdiff{E}{\lambda^\mu}}.
\end{equation}
These are the expectations of the operators coupled to $\lambda^\mu$.
As these quantities are functions of the parameters $\lambda^\mu$, they provide another coordinate system for the manifold of control parameters.
Any linear combination of these parameters would work equally well.

The optimal unbiased estimators for the $\tlambda^\mu$ are
\begin{equation}\label{eq:dualest}
  \tlh^\mu = \CO_\mu = - \pdiff{E}{\lambda^\mu},
\end{equation}
\ie the operator to which $\lambda^\mu$ couples in the Hamiltonian.
These estimators are unbiased, by comparison to \eqref{eq:dual}.
The Fisher information in the original $\lambda^\mu$ coordinate system can be computed from the second derivatives of the logarithm of the partition function.
Due to the fluctuation-dissipation theorem, this quantity is also equal to the covariance of the operators and the susceptibilities (up to factors of $\kt$):
\begin{equation*}
  \frac{\partial^2 \ln \mathcal{Z}}{\partial \lambda^\mu \partial \lambda^\nu} = \fisher_{\mu\nu} = \beta^2 \av{\delta\CO_\mu \delta\CO_\nu} = \beta \pdiff{\av{\CO_\mu}}{\lambda^\nu}.
\end{equation*}
Noting that the third quantity is $\beta^2 \covesth_{\mu\nu}$, the covariance of the estimators \eqref{eq:dualest}, and the last one is $\pdiff{\tlambda^\mu}{\lambda^\nu}$, the Jacobian matrix for the coordinate change \eqref{eq:dual}, the Fisher information in the new $\tlambda^\mu$ coordinate system is given by
\begin{equation*}
\begin{aligned}
  \fisherh_{\mu\nu} &= \pdiff{\lambda^\rho}{\tlambda^\mu} \fisher_{\rho\sigma} \pdiff{\lambda^\sigma}{\tlambda^\nu}
    = \brk{\beta\fisher\inv \cdot \fisher \cdot \beta\fisher\inv}_{\mu\nu}
    \\&= \beta^2 \fisher\inv_{\mu\nu}
    = \covesth\inv_{\mu\nu} = \preccovh_{\mu\nu}.
\end{aligned}
\end{equation*}
Therefore, these estimators saturate the Cram\'er-Rao bound.


\section{Example systems: further details}\label{sec:examples_details}

Here we provide further details for the example systems that were introduced in the \autoref{sec:examples}.


\subsection{Heavily over-damped harmonic oscillator}\label{sec:damposc_details}

Here we will provide further details for the heavily damped harmonic oscillator presented in the \autoref{sec:damposc}.
This system is described by the following Langevin equation:
\begin{equation}\label{eq:damposcs}
  \zeta \dot{x} = \kappa(\lambda(t) - x) + \sqrt{2\zeta\kt} \xi(t),
\end{equation}
where $\xi(t)$ is a Gaussian process with $\av{\xi(t)}=0$ and $\av{\xi(t)\xi(t')} = \delta(t-t')$.
The energy of this system is given by:
\begin{equation}\label{eq:dampescenergy}
  E = \frac{1}{2} \kappa (x-\lambda)^2,
\end{equation}
with the kinetic energy being neglected in the overdamped limit.

Introducing the quantities $\tau = \frac{\zeta}{\kappa}$ and $\sigma = \sqrt{\frac{\kt}{\kappa}}$, the solution of \eqref{eq:damposcs} is
\begin{equation}\label{eq:damposcsol}
  x(t) = \int_0^\infty \!\!\frac{\dr t'}{\tau} \, \e^{-t'/\tau} \prn{ \lambda(t-t') + \sqrt{2\tau}\sigma \xi(t-t') }\!.
\end{equation}
Therefore, $x(t)$ is a Gaussian process with
\begin{equation}\label{eq:damposcmeanfull}
\begin{aligned}
  \av{x(t)} &= \int_0^\infty \!\!\frac{\dr t'}{\tau} \, \e^{-t'/\tau} \lambda(t-t') \equiv \mu(t),
  \\
  \av{\delta x(t) \delta x(t')} &= \sigma^2 \e^{-\abs{t-t'}/\tau}.
\end{aligned}
\end{equation}
We can express $\mu(t)$ as a derivative expansion by Taylor expanding $\lambda(t-t')$ in $t'$ to find
\begin{equation}\label{eq:damposcseries}
  \mu(t) = \sum_{n=0}^{\infty} \brk{-\tau\diff{}{t}}^n \lambda(t)
   \approx \lambda(t),
\end{equation}
where the approximation is valid when the timescale over which $\lambda(t)$ varies is much larger than $\tau$.
More precisely, looking at the Fourier transforms
\begin{equation}\label{eq:damposcFT}
  {\mu}(\omega) = \frac{{\lambda}(\omega)}{1-\ir\,\omega\tau}
    = \sum_{n=0}^{\infty} \prn{\ir\,\omega\tau}^n {\lambda}(\omega).
\end{equation}
Defining $\omega_\text{max}$ as the largest $\omega$ for which $\widetilde{\lambda}(\omega)$ is significantly nonzero, when $\omega_\text{max}\tau \ll 1$ as in \eqref{eq:valid}, we are justified in neglecting all of the higher order terms in this series.

In this regime, the optimal unbiased estimator of $\lambda(t)$ is
\begin{equation}\label{eq:damposcest}
  \hat{\lambda}(t) = x(t),
  \quad
  \varest = \sigma^2,
  \quad
  \precest = \frac{1}{\sigma^2}.
\end{equation}

The excess power is given by
\begin{equation}\label{eq:damposcpowfull}
\begin{aligned}
  \pow &= \ld(t) \av{\pdiff{E(t)}{\lambda}} \\
   &= \kappa \ld(t) \prn{\lambda(t)-\mu(t)} \\
   &= \kappa \ld(t) \int_0^\infty \!\!\dr t' \, \e^{-t'/\tau} \ld(t-t') \\
   &= \zeta \ld(t) \sum_{n=0}^{\infty} (-\tau)^n \diff[n+1]{\lambda(t)}{t}.
\end{aligned}
\end{equation}
By the same logic used when making the approximation $\mu(t) \approx \lambda(t)$, in the regime where timescale over which $\lambda(t)$ varies is much larger than $\tau$ we can neglect the higher order terms to find
\begin{equation}\label{eq:damposxpowmet}
  \pow \approx \zeta \ld(t)^2.
\end{equation}

To calculate $\tmin$, we need to look at the Fokker-Planck equation associated with \eqref{eq:damposcs}:
\begin{equation}\label{eq:damposcFP}
  \pdiff{P(x,t)}{t} = \pdiff{}{x}\brk{\frac{(x-\lambda)P(x,t)}{\tau}} + \brk{\frac{\sigma^2}{\tau}} \pdiff{^2 P(x,t)}{x^2}.
\end{equation}
The eigenfunctions of the operator on the right are
\begin{equation}\label{eq:damposceig}
\begin{aligned}
  \evl^n(x) &= \frac{ \e^{-(x-\lambda)^2/2\sigma^2} }{ \sqrt{2^{n+1}\pi n!}\sigma } H_n\prn{\frac{x-\lambda}{\sqrt{2}\sigma}}, \\
  \evr^n(x) &= \frac{1}{\sqrt{2^n n!}} H_n\prn{\frac{x-\lambda}{\sqrt{2}\sigma}}, \\
  \tau_n &= \frac{\tau}{n},
\end{aligned}
\end{equation}
where $H_n(x)$ are Hermite polynomials.

The ``force'' dual to $\lambda$ is given by
\begin{equation}\label{eq:damposcforce}
\begin{aligned}
  \delta\phi(x) &= -\beta\prn{\pdiff{E}{\lambda} - \av{\pdiff{E}{\lambda}}} \\
    &= \frac{x-\lambda}{\sigma^2}
    = \frac{\evr^1(x)}{\sigma}.
\end{aligned}
\end{equation}
Therefore, the coupling between $\lambda$ and the eigenmodes is
\begin{equation}\label{eq:damposccouple}
  \int\!\!\dr x \,  \evl^n(x) \delta\phi(x) = \frac{\delta_{n,1}}{\sigma}.
\end{equation}
As $\lambda$ only couples to the first mode, we have $\tmin=\tau$.


\subsection{Ising ring}\label{sec:ising_details}

Here we provide the details underlying the calculations summarized in the \autoref{sec:ising} for the Ising model 
and extend them to the construction of estimators and varying $J$.
The Hamiltonian of this system is
\begin{equation*}
  H = - h \sum_n \sigma_n - J \sum_n \sigma_n \sigma_{n+1}.
\end{equation*}
The sender of the signal will vary $h$ and $J$ and the receiver will observe the spins.
We will perform all computations at the instant when we pass through $h=0$, although $\dot{h}$ is not necessarily zero.
It will be convenient to use the notation
\begin{equation*}
  \theta = \tanh \beta h,
  \qquad
  \gamma = \tanh 2\beta J,
  \qquad
  \xi = \tanh \beta J.
\end{equation*}
This system undergoes Glauber dynamics \cite{Glauber1963}, \ie the rate at which spin $n$ flips is given by
\begin{equation*}
  w_n = \frac{\alpha}{2}  \brk{ 1 - \theta\, \sigma_n + \frac{\gamma}{2}  (\theta-\sigma_n) (\sigma_{n-1}+\sigma_{n+1}) },
\end{equation*}
where $\alpha$ is an overall rate.
These dynamics satisfy detailed balance.

When $h$ and $\theta$ are zero, the relevant correlation functions have been computed \cite{Glauber1963,Mayer2004}:
\begin{equation}\label{eq:isingphicorr}
\begin{gathered}
\begin{aligned}
  \av{ \delta\phi_h(0) \delta\phi_h(t) }
    &= N \beta^2 \e^{2J} \prn{\frac{1-\xi^N}{1+\xi^N}} \e^{-\alpha(1-\gamma)t}  , \\
  \av{ \delta\phi_J(0) \delta\phi_J(t) }
    &= \beta^2 \sum_{m=0}^{N-1} \frac{2(1-\gamma^2)\sin^2q_m}{\nu(q_m)^2} \, \e^{-2\alpha\nu(q_m)t} ,
\end{aligned}\\
\begin{aligned}
    \text{where} \quad
    \nu(q) &= 1-\gamma\cos q \\
    \text{and} \quad
    q_m &= \frac{2\pi}{N} \prn{ m + \frac{1}{2} }.
\end{aligned}
\end{gathered}
\end{equation}
All off-diagonal components vanish at $h=0$ by symmetry.
This is sufficient to compute the dissipation rate when $\dot{h}$ and $\dot{J}$ are nonzero at the instant when $h=0$, but not for computing finite distances away from that line.

In the large $N$ limit, the friction tensor and Fisher information are given by
\begin{equation*}
\begin{aligned}
  \thermmet_{hh} &= \frac{N \beta \e^{2\beta J}}{\alpha(1-\gamma)}, &
  \thermmet_{JJ} &= \frac{N \beta(1+\xi^2)}{\alpha} , \\
  \fisher_{hh} &= N \beta^2\e^{2\beta J}, &
  \fisher_{JJ} &= N \beta^2(1-\xi^2) . \\
\end{aligned}
\end{equation*}

The timescales of the eigenmodes were given in \cite{Glauber1963}.
They are
\begin{equation*}
  \frac{1}{\tau_a} = \sum_r \alpha \nu(q_{r}).
\end{equation*}
These can be loosely thought of multi-particle states, with each particle a plane-wave superposition of single spin flips, $q_{r}$ their (distinct) momenta and $\alpha\nu(q)$ the dispersion relation.
The shortest of these timescales is $\frac{1}{\alpha N}$.
If we used this as $\tmin$ we would have a very loose bound.
Instead, if we look at the timescales that appear in the correlation functions \eqref{eq:isingphicorr}, we see that these parameters only couple to a subset of the eigenmodes.
The magnetic field, $h$, only couples to the mode with one particle of zero momentum.
The ferromagnetic interaction, $J$, couples to modes with two particles of equal and opposite momenta.
Finding the shortest of these timescales gives
\begin{equation*}
  \tmin = \frac{1}{2\alpha(1+\abs{\gamma})} = \frac{\e^{-2\beta\abs{J}}\cosh 2\beta J}{2\alpha}.
\end{equation*}

It is difficult to construct unbiased estimators for $h$ and $J$.
Instead, in the large $N$ limit, we will use estimators for the following dual parameters (see \eqref{eq:dual})
\begin{equation*}
\begin{gathered}
\begin{aligned}
  \tlambda^1 &= \frac{\theta(1+\xi)}{(1+\zeta)(1-\xi)}, &
  \tlh^1 &= \frac{\sum_n\sigma_n}{N}, \\
  \tlambda^2 &= \frac{\xi + (1-\xi)\prn{\frac{\theta^2}{1+\zeta}-\frac{\zeta}{2}}}{1+(1-\xi)\frac{\zeta}{2}}, &
  \tlh^2 &= \frac{\sum_n\sigma_n\sigma_{n+1}}{N},
\end{aligned}\\
  \text{where} \quad \zeta = \sqrt{1+\frac{4\theta^2\xi}{(1-\xi)^2}} - 1.
\end{gathered}
\end{equation*}
For small $h$, $\tlambda^1 \approx \e^{2\beta J}\tanh \beta h$ and $\tlambda^2 \approx \tanh \beta J$.
When $h=0$, the covariance of these estimators is
\begin{equation*}
  \covest^{11} = \frac{\e^{2\beta J}}{N},
  \qquad
  \covest^{22} = \frac{\sech^2\!\!\beta J}{N},
\end{equation*}
and the friction tensor in these coordinates is
\begin{equation*}
\begin{aligned}
  \thermmet_{11} &= \frac{N(\kt)\cosh2\beta J}{\alpha}, \\
  \thermmet_{22} &= \frac{N(\kt)\cosh2\beta J\cosh^2\!\!\beta J}{\alpha}.
\end{aligned}
\end{equation*}

These quantities can be used to investigate the power--precision--speed bound \eqref{eq:matrixbound}:
\begin{equation}\label{eq:isingpowbnd}
  \frac{\tr\!\prn{\preccovb\spdb}}{\pow}
    = \frac{\alpha\beta}{\cosh2\beta J} \frac{\e^{-2J}(\tld^1)^2 + \cosh^2\!\!\beta J(\tld^2)^2}{(\tld^1)^2 + \cosh^2\!\!\beta J(\tld^2)^2}.
\end{equation}
This is maximized either when $\tld^1=0$ or when $\tld^2=0$, leading to 
\begin{equation}\label{eq:isingpowbndmin}
  \frac{\tr\!\prn{\preccovb\spdb}}{\pow}
    \leq \frac{\alpha\beta}{\e^{2\beta[J]_-} \cosh2\beta J}
    = \frac{\frac{1}{2} \e^{-2\beta[J]_+}}{(\kt) \, \tmin},
\end{equation}
We can also compute the mean power use,
\begin{multline}\label{eq:isingpowavbnd}
    \beta\av{\pow}\tr(\covestb \mspdb\inv) =\\
      \frac{ \prn{\mspd^{11}+\mspd^{22}\cosh^2\!\!\beta J} \prn{\mspd^{11}+\mspd^{22}\e^{2\beta J}\cosh^2\!\!\beta J} }{\alpha\sech 2\beta J \cosh^2\!\!\beta J\prn{\mspd^{11}\mspd^{22}-\mspd^{12}\mspd^{21}}}.
\end{multline}
This is minimized when $\mspd^{12}=\mspd^{21}=0$ and $\frac{\mspd^{11}}{\mspd^{22}} = \e^{\beta J} \cosh^2\!\!\beta J$,
leading to 
\begin{equation}\label{eq:isingpowavbndmin}
\begin{aligned}
    \beta\av{\pow}\tr(\covestb \mspdb\inv)
          &\geq \frac{\cosh2\beta J(1+\e^\beta J)^2}{\alpha} \\
          &= \frac{\e^{2\beta\abs{J}}(1+\e^{\beta J})^2}{2} \cdot 4\tmin.
\end{aligned}
\end{equation}

So our bounds, \eqref{eq:matrixbound} and \eqref{eq:ppsbound}, are tighter at weak coupling/high temperature but very loose at strong coupling/low temperature.
However, the looseness of this bound does not scale with the size of the system, $N$, as one might have worried based on the proof of the inequality \eqref{eq:detbalmet}.
This is another illustration of the importance of computing the coupling of the parameters to the eigenmodes when computing $\tmin$.

If we only allow $h$ to vary and only estimate $\tlambda^1$, we only include the mode that $h$ couples to when calculating $\tmin$:
\begin{equation*}
  \tmin = \frac{\e^{2\beta J} \cosh 2\beta J}{\alpha}.
\end{equation*}
In this case, we have
\begin{equation*}
  \pow = \frac{N(\kt)\, \tld^2 \cosh 2\beta J}{\alpha},
  \qquad
  \precesth = N\e^{-2\beta J}.
\end{equation*}
We can then investigate the bound in \eqref{eq:scalarbound}
\begin{equation*}
  \frac{\precest \, \spd }{\pow}
   =  \frac{\alpha} {(\kt)\,\e^{2\beta J} \cosh 2\beta J}
   = \frac{1}{(\kt)\,\tmin},
\end{equation*}
which saturates the bound.


\subsection{A non-equilibrium four state receptor}\label{sec:receptor}

\begin{figure}
  \begin{center}
  \includegraphics[width=0.6\linewidth]{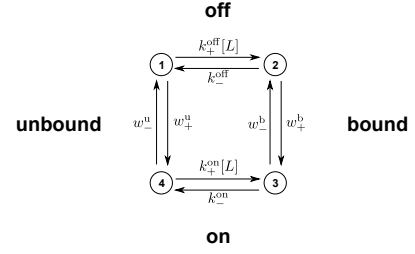}
  \end{center}
  \caption{Single four-state receptor, adapted from \cite[Fig.1a]{Skoge2013}.
  Arrows are labelled with transition rates.
  Ligand concentration is indicated by $[L]$.
  In states 2 and 3, the receptor is bound to a ligand molecule.
  In states 3 and 4, the receptor is in an activated state.}\label{fig:receptor}
\end{figure}

As an example of a system that does not satisfy detailed balance, we will study the four-state model of a single receptor in \cite[Fig.1a]{Skoge2013}, as shown in \autoref{fig:receptor}.
we will be treating this system as a Markov process rather than a chemical reaction.
The control parameter will be the external ligand concentration, so the physical interpretation of the source of the work is unclear.

Following \cite{Skoge2013}, we will assume that $k_+\on = k_+\off = k_+$.
We will use the following notation
\begin{equation}\label{eq:recparam}
  k_- = \sqrt{k_-\off k_-\on},
  \quad
  \kappa = \ln \frac{k_-\off}{k_-\on},
  \quad
  \lambda = \ln \frac{k_+ [L]}{k_-}.
\end{equation}
Activation and inactivation are described by heat bath kinetics
\begin{equation}\label{eq:recw}
  w^\text{u/b}_\pm = \frac{\alpha}{1+\e^{\pm\beta\Delta\CE^\text{u/b}}},
  \qquad
  \alpha \ll k_-.
\end{equation}
The thermodynamic driving force, which is a measure of the violation of detailed balance, is given by
\begin{equation}\label{eq:recnoneq}
  \e^{-\gamma} = \frac{k_+\off w_+^\text{b} k_-\on w_-^\text{u}}{k_-\off w_-^\text{b} k_+\on w_+^\text{u}},
  \; \implies \;
  \beta(\Delta\CE^\text{u}-\Delta\CE^\text{b}) = \kappa - \gamma.
\end{equation}
Following \cite{Skoge2013}, we make the choice
\begin{equation}\label{eq:recassume}
  \Delta\CE^\text{u}=-\Delta\CE^\text{b} = \frac{\kt}{2} \, ( \kappa - \gamma),
\end{equation}

At the point $\lambda=0$, the eigenvectors and eigenvalues of the matrix $\fundh$ that appears in \eqref{eq:fundsymev} can be computed perturbatively in the small parameter $\frac{\alpha}{k_-}$.

\begin{equation}\label{eq:rectime}
\begin{aligned}
  \tauh_1 &= \frac{1}{k_-(1+\e^{\kappa/2})} + \CO\!\prn{\frac{\alpha}{k_-^2}}, \\
  \tauh_2 &= \frac{1}{k_-(1+\e^{-\kappa/2})} + \CO\!\prn{\frac{\alpha}{k_-^2}}, \\
  \tauh_3 &= \frac{\cosh\prn{\frac{\kappa-\gamma}{4}}\cosh\prn{\frac{\kappa}{4}}}{\alpha\cosh\prn{\frac{\gamma}{4}}}
  + \CO\!\prn{k_-\inv}.
\end{aligned}
\end{equation}
The couplings to the parameter $\lambda$ are
\begin{equation}\label{eq:reccouple}
\begin{aligned}
  \evlhb^1 \cdt \delta\phib_\lambda &= \frac{\sech\prn{\frac{\kappa}{4}}}{2\sqrt{2}} + \CO\!\prn{\frac{\alpha}{k_-}} ,\\
  \evlhb^2 \cdt \delta\phib_\lambda &= \frac{\sech\prn{\frac{\kappa}{4}}}{2\sqrt{2}} + \CO\!\prn{\frac{\alpha}{k_-}} ,\\
  \evlhb^3 \cdt \delta\phib_\lambda &= \frac{\tanh\prn{\frac{\kappa}{4}}-\tanh\prn{\frac{\gamma}{4}}}{2}
  + \CO\!\prn{\frac{\alpha}{k_-}}.
\end{aligned}
\end{equation}
As in the case of the \hyperref[sec:ising_details]{Ising ring}, these are sufficient to compute the dissipation rate when $\ld$ is nonzero at the instant when $\lambda=0$, but not at a finite distance from that point.

We can use these to compute the friction tensor and Fisher information:
\begin{equation*}
\begin{aligned}
  \thermmet_{\lambda\lambda} &= \frac{(\kt)\cosh\prn{\frac{\kappa-\gamma}{4}}\sinh^2\prn{\frac{\kappa-\gamma}{4}}}{16\alpha\cosh^3\prn{\frac{\gamma}{4}}\cosh\prn{\frac{\kappa}{4}}}
              + \CO\!\prn{k_-\inv}, \\
  \fisher_{\lambda\lambda} &= \frac{\cosh\prn{\frac{\kappa-2\gamma}{4}}}{16\cosh^2\prn{\frac{\gamma}{4}}\cosh\prn{\frac{\kappa}{4}}} + \CO\!\prn{\frac{\alpha}{k_-}}.
\end{aligned}
\end{equation*}
From \eqref{eq:rectime}, we see that $\tminh=\frac{1}{k_-(1+\e^{\abs{\kappa}/2})}$, so generically the bound \eqref{eq:metgeneig} is loose by a factor $\CO(k_-/\alpha)$.
This is because the system has one very long timescale, $\tauh_3$, and two very short timescales, $\tauh_1$ and $\tauh_2$, all of which couple to the parameter $\lambda$.%
\footnote{This appears to still be true away from the special point $\lambda=0$, but the algebra is much more complicated.}
At the special parameter values $\gamma=\kappa$, the parameter does not couple to the slowest timescale and the bound is only loose by a factor of $\frac{1}{2}(1+\e^{\abs{\kappa}/2})$.
However, this is the point where $\Delta\CE^\text{u}=\Delta\CE^\text{b}$ (see \eqref{eq:recnoneq}), so the relative occupation of the on/off states contains no information about $[L]$.

Defining an observable $\sigma$ that is +1 in the on states and -1 in the off states, we can use it as an unbiased estimator for the dual parameter $\tlambda$ (see \eqref{eq:dual})
\begin{equation*}
\begin{aligned}
  \tlambda &= \frac{ \sinh\prn{\frac{\kappa-\gamma}{4}}\sinh\lambda }{ \cosh\prn{\frac{\kappa-\gamma}{4}}\cosh\lambda +\cosh\prn{\frac{\kappa+\gamma}{4}} },
  \\
  \varesth &= 1-\tlambda{}^2 + \CO\!\prn{\frac{\alpha}{k_-}},\\
  \precesth &= \frac{1}{1-\tlambda^2} + \CO\!\prn{\frac{\alpha}{k_-}}.
\end{aligned}
\end{equation*}
This estimator is more sensible when we consider many independent receptors.
This will scale down the estimator variance by $1/N$ and scale up the power consumption by $N$.
The parameter will only couple to the eigenmodes with timescales $\tauh_{1,2,3}$, so $\tminh$ will be unchanged.

\begin{figure}
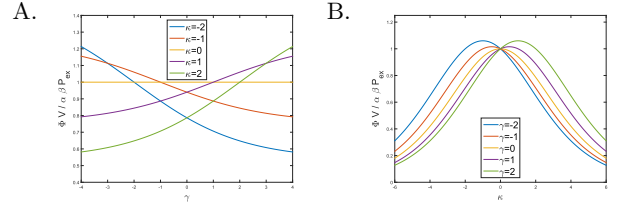

  \begin{center}
  \begin{myenuma}
    \item\aligntop{\includegraphics[width=0.42\linewidth]{recpowplotg.eps}}\label{fig:recpowneq}
    \item\aligntop{\includegraphics[width=0.42\linewidth]{recpowplotk.eps}}\label{fig:recpowbind}
  \end{myenuma}
  \end{center}
  \caption{Power-precision-speed trade-off as a function of \ref{fig:recpowneq}. non-equilibrium parameter, $\gamma$, and \ref{fig:recpowbind}. log unbinding ratio, $\kappa$.}\label{fig:recpower}
\end{figure}

With this parameterization, the friction tensor is
\begin{equation*}
\begin{aligned}
  \thermmet_{\tlambda\tlambda} &= \frac{(\kt) \cosh\prn{\frac{\kappa-\gamma}{4}} \cosh\prn{\frac{\kappa}{4}} }{ \alpha \cosh\prn{\frac{\gamma}{4}} }
              + \CO\!\prn{k_-\inv},\\
\end{aligned}
\end{equation*}
We can then compute all quantities in the bound \eqref{eq:ppsboundgen} to find
\begin{equation}\label{eq:recbnd}
\begin{aligned}
  \frac{\precest \, \spd}{\pow} &= \frac{ \alpha\beta \cosh\prn{\frac{\gamma}{4}} }{ \cosh\prn{\frac{\kappa-\gamma}{4}} \cosh\prn{\frac{\kappa}{4}} }
    + \CO\!\prn{k_-\inv}\\
    &\ll \frac{1}{(\kt) \, \tminh}.
\end{aligned}
\end{equation}
We see that the thermodynamic driving force, $\gamma$,  has no effect on the right hand side of this bound.
What it does is allow us to get closer to the bound by reducing the timescale of the slowest mode,
as shown in \autoref{fig:recpower}.

When $\abs{\kappa}$ is large, the coupling to the two fastest modes becomes very small, as seen in \eqref{eq:reccouple}.
In this regime, one could neglect these two modes and set $\tminh = \tauh_3$.
In this case, the bound \eqref{eq:ppsboundgen} would be saturated.

We see in \autoref{fig:recpower}, that at fixed $\kappa$ it is always helpful to increase $\gamma$ (with the same sign as $\kappa$).
However, at fixed $\gamma$ this is minimized at $\kappa = \frac{\gamma}{2}$.
Note that we made a number of assumptions in the choice of parameters in \eqref{eq:recparam}, \eqref{eq:recw} and \eqref{eq:recassume}, following \cite{Skoge2013}, as well as working in the regime where the ligand concentration varies slowly compared to the intrinsic timescales of the system and neglecting receptor cooperativity.
In addition, we only computed excess power, neglecting the power consumed to maintain the non-equilibrium steady state.
Furthermore, the levels of ATP, \etc are needed to support many other cellular processes.
Therefore, the special value $\gamma=2\kappa$ should not be taken too seriously as a prediction.
Note that for $\kappa<\gamma$, the occupation of the on states is anticorrelated with $[L]$.

\end{appendices}
\bibliography{pps_paper-minimal}

\providecommand{\href}[2]{#2}\begingroup\raggedright\begin{thebibliography}{10}

\bibitem{Thorpe1996speed}
S.~Thorpe, D.~Fize, and C.~Marlot, ``{Speed of processing in the human visual
  system.},'' \href{http://dx.doi.org/10.1038/381520a0}{{\em Nature} {\bf 381}
  (jun, 1996)  520--2}.

\bibitem{sterling2015principles}
P.~Sterling and S.~Laughlin, {\em Principles of neural design}.
\newblock MIT Press, 2015.

\bibitem{Shannon1948infotheory}
C.~E. Shannon, ``{A Mathematical Theory of Communication},''
  \href{http://dx.doi.org/10.1002/j.1538-7305.1948.tb01338.x}{{\em Bell System
  Technical Journal} {\bf 27} (jan, 1948)  379--423, 623--656}.

\bibitem{cover2012infotheory}
T.~M. Cover and J.~A. Thomas, {\em Elements of information theory}.
\newblock John Wiley \& Sons, 2012.

\bibitem{Gastpar2003code}
M.~Gastpar, B.~Rimoldi, and M.~Vetterli, ``{To code, or not to code: lossy
  source-channel communication revisited},''
  \href{http://dx.doi.org/10.1109/TIT.2003.810631}{{\em IEEE Transactions on
  Information Theory} {\bf 49} (May, 2003)  1147--1158}.

\bibitem{Polyanskiy2010dispersion}
Y.~Polyanskiy, H.~V. Poor, and S.~Verdu, ``{Channel Coding Rate in the Finite
  Blocklength Regime},'' \href{http://dx.doi.org/10.1109/TIT.2010.2043769}{{\em
  IEEE Transactions on Information Theory} {\bf 56} (May, 2010)  2307--2359}.

\bibitem{Schmiedl2007finitetime}
T.~Schmiedl and U.~Seifert, ``{Optimal finite-time processes in stochastic
  thermodynamics},''
  \href{http://dx.doi.org/10.1103/PhysRevLett.98.108301}{{\em Phys. Rev. Lett.}
  {\bf 98} (Mar., 2007)  108301},
  \href{http://arxiv.org/abs/cond-mat/0701554}{{\tt arXiv:cond-mat/0701554
  [cond-mat]}}.

\bibitem{Then2008finitetime}
H.~Then and A.~Engel, ``{Computing the optimal protocol for finite-time
  processes in stochastic thermodynamics},''
  \href{http://dx.doi.org/10.1103/PhysRevE.77.041105}{{\em Phys. Rev. E} {\bf
  77} (Apr., 2008)  041105}, \href{http://arxiv.org/abs/0710.3297}{{\tt
  arXiv:0710.3297 [cond-mat.stat-mech]}}.

\bibitem{Landauer1961irrev}
R.~Landauer, ``{Irreversibility and Heat Generation in the Computing
  Process},'' \href{http://dx.doi.org/10.1147/rd.53.0183}{{\em IBM Journal of
  Research and Development} {\bf 5} (July, 1961)  183--191}.

\bibitem{Bennett1973irrev}
C.~H. Bennett, ``{Logical Reversibility of Computation},''
  \href{http://dx.doi.org/10.1147/rd.176.0525}{{\em IBM Journal of Research and
  Development} {\bf 17} (Nov., 1973)  525--532}.

\bibitem{Bennett1982review}
C.~H. Bennett, ``{The thermodynamics of computation—-a review},''
  \href{http://dx.doi.org/10.1007/BF02084158}{{\em International Journal of
  Theoretical Physics} {\bf 21} (Dec., 1982)  905--940}.

\bibitem{Toffoli1980reversible}
T.~Toffoli, \href{http://dx.doi.org/10.1007/3-540-10003-2_104}{``Reversible
  computing,''} in {\em Automata, Languages and Programming}, J.~de~Bakker and
  J.~van Leeuwen, eds., vol.~85 of {\em Lecture Notes in Computer Science},
  pp.~632--644.
\newblock Springer Berlin Heidelberg, 1980.

\bibitem{Hopfield1974}
J.~J. Hopfield, ``{Kinetic Proofreading: A New Mechanism for Reducing Errors in
  Biosynthetic Processes Requiring High Specificity},''
  \href{http://dx.doi.org/10.1073/pnas.71.10.4135}{{\em Proc. Natl. Acad. Sci.
  U.S.A.} {\bf 71} (Oct., 1974)  4135--4139}.

\bibitem{Savageau1979}
M.~A. Savageau and R.~R. Freter, ``{On the evolution of accuracy and cost of
  proofreading tRNA aminoacylation.},''
  \href{http://dx.doi.org/10.1073/pnas.76.9.4507}{{\em Proc. Natl. Acad. Sci.
  U.S.A.} {\bf 76} (sep, 1979)  4507--10}.

\bibitem{Savageau1979a}
M.~A. Savageau and R.~R. Freter, ``{Energy cost of proofreading to increase
  fidelity of transfer ribonucleic acid aminoacylation},''
  \href{http://dx.doi.org/10.1021/bi00583a008}{{\em Biochemistry} {\bf 18}
  (Aug., 1979)  3486--3493}.

\bibitem{Freter1980}
R.~R. Freter and M.~A. Savageau, ``{Proofreading systems of multiple stages for
  improved accuracy of biological discrimination},''
  \href{http://dx.doi.org/10.1016/0022-5193(80)90284-2}{{\em Journal of
  Theoretical Biology} {\bf 85} (July, 1980)  99--123}.

\bibitem{Ehrenberg1980}
M.~Ehrenberg and C.~Blomberg, ``{Thermodynamic constraints on kinetic
  proofreading in biosynthetic pathways.},''
  \href{http://dx.doi.org/10.1016/S0006-3495(80)85063-6}{{\em Biophysical
  journal} {\bf 31} (Sept., 1980)  333--58}.

\bibitem{Blomberg1981}
C.~Blomberg and M.~Ehrenberg, ``{Energy considerations for kinetic proofreading
  in biosynthesis},''
  \href{http://dx.doi.org/10.1016/0022-5193(81)90242-3}{{\em Journal of
  Theoretical Biology} {\bf 88} (Feb., 1981)  631--670}.

\bibitem{Savageau1981}
M.~A. Savageau and D.~S. Lapointe, ``{Optimization of kinetic proofreading: A
  general method for derivation of the constraint relations and an exploration
  of a specific case},''
  \href{http://dx.doi.org/10.1016/0022-5193(81)90062-X}{{\em Journal of
  Theoretical Biology} {\bf 93} (Nov., 1981)  157--177}.

\bibitem{Qian2006}
H.~Qian, ``{Reducing intrinsic biochemical noise in cells and its thermodynamic
  limit.},'' \href{http://dx.doi.org/10.1016/j.jmb.2006.07.068}{{\em Journal of
  molecular biology} {\bf 362} (Sept., 2006)  387--92}.

\bibitem{Murugan2014}
A.~Murugan, D.~A. Huse, and S.~Leibler, ``{Discriminatory Proofreading Regimes
  in Nonequilibrium Systems},''
  \href{http://dx.doi.org/10.1103/PhysRevX.4.021016}{{\em Phy. Rev. X} {\bf 4}
  (Apr., 2014)  021016}, \href{http://arxiv.org/abs/1312.2286}{{\tt
  arXiv:1312.2286 [cond-mat.stat-mech]}}.

\bibitem{Murugan2012}
A.~Murugan, D.~A. Huse, and S.~Leibler, ``{Speed, dissipation, and error in
  kinetic proofreading.},''
  \href{http://dx.doi.org/10.1073/pnas.1119911109}{{\em Proc. Natl. Acad. Sci.
  U.S.A.} {\bf 109} (July, 2012)  12034--9}.

\bibitem{Endres2009}
R.~G. Endres and N.~S. Wingreen, ``{Maximum likelihood and the single
  receptor},'' \href{http://dx.doi.org/10.1103/PhysRevLett.103.158101}{{\em
  Phys. Rev. Lett.} {\bf 103} (Oct., 2009)  158101},
  \href{http://arxiv.org/abs/0909.4710}{{\tt arXiv:0909.4710 [q-bio.SC]}}.

\bibitem{Mehta2012}
P.~Mehta and D.~J. Schwab, ``{Energetic costs of cellular computation.},''
  \href{http://dx.doi.org/10.1073/pnas.1207814109}{{\em Proc. Natl. Acad. Sci.
  U.S.A.} {\bf 109} (Oct., 2012)  17978--82},
  \href{http://arxiv.org/abs/1203.5426}{{\tt arXiv:1203.5426 [q-bio.MN]}}.

\bibitem{Lang2014}
A.~H. Lang, C.~K. Fisher, T.~Mora, and P.~Mehta, ``{Thermodynamics of
  statistical inference by cells},''
  \href{http://dx.doi.org/10.1103/PhysRevLett.113.148103}{{\em Phys. Rev.
  Lett.} {\bf 113} (May, 2014)  148103},
  \href{http://arxiv.org/abs/1405.4001}{{\tt arXiv:1405.4001
  [physics.bio-ph]}}.

\bibitem{Barato2014}
A.~C. Barato, D.~Hartich, and U.~Seifert, ``{Efficiency of cellular information
  processing},'' \href{http://dx.doi.org/10.1088/1367-2630/16/10/103024}{{\em
  New Journal of Physics} {\bf 16} (may, 2014)  },
  \href{http://arxiv.org/abs/1405.7241}{{\tt arXiv:1405.7241}}.

\bibitem{Govern2014diss}
C.~C. Govern and P.~R. ten Wolde, ``{Energy Dissipation and Noise Correlations
  in Biochemical Sensing},''
  \href{http://dx.doi.org/10.1103/PhysRevLett.113.258102}{{\em Phys. Rev.
  Lett.} {\bf 113} (Dec., 2014)  258102}.

\bibitem{Govern2014resource}
C.~C. Govern and P.~R. {Ten Wolde}, ``{Optimal resource allocation in cellular
  sensing systems.},'' \href{http://dx.doi.org/10.1073/pnas.1411524111}{{\em
  Proc. Natl. Acad. Sci. U.S.A.} {\bf 111} (Nov., 2014)  17486--17491}.

\bibitem{Sartori2014}
P.~Sartori, L.~Granger, C.~Lee, and J.~Horowitz, ``{Thermodynamic costs of
  information processing in sensory adaptation.},''
  \href{http://dx.doi.org/10.1371/journal.pcbi.1003974}{{\em PLoS computational
  biology} {\bf 10} (Dec., 2014)  e1003974},
  \href{http://arxiv.org/abs/1404.1027}{{\tt arXiv:1404.1027
  [cond-mat.stat-mech]}}.

\bibitem{Lan2012}
G.~Lan, P.~Sartori, S.~Neumann, V.~Sourjik, and Y.~Tu, ``{The
  energy-speed-accuracy tradeoff in sensory adaptation.},''
  \href{http://dx.doi.org/10.1038/nphys2276}{{\em Nature physics} {\bf 8} (May,
  2012)  422--428}.

\bibitem{Dally2004}
W.~J. Dally and B.~P. Towles, {\em {Principles and Practices of Interconnection
  Networks}}.
\newblock Elsevier, 2004.

\bibitem{Sivak2012metric}
D.~A. Sivak and G.~E. Crooks, ``{Thermodynamic Metrics and Optimal Paths},''
  \href{http://dx.doi.org/10.1103/PhysRevLett.108.190602}{{\em Phys. Rev.
  Lett.} {\bf 108} (May, 2012)  190602},
  \href{http://arxiv.org/abs/1201.4166}{{\tt arXiv:1201.4166
  [cond-mat.stat-mech]}}.

\bibitem{Zulkowski2012geo}
P.~R. Zulkowski, D.~A. Sivak, G.~E. Crooks, and M.~R. Deweese, ``{Geometry of
  thermodynamic control},''
  \href{http://dx.doi.org/10.1103/PhysRevE.86.041148}{{\em Phys. Rev. E} {\bf
  86} (Oct., 2012)  041148}, \href{http://arxiv.org/abs/1208.4553}{{\tt
  arXiv:1208.4553 [cond-mat.stat-mech]}}.

\bibitem{Zulkowski2013noneq}
P.~R. Zulkowski, D.~A. Sivak, and M.~R. DeWeese, ``{Optimal control of
  transitions between nonequilibrium steady states.},''
  \href{http://dx.doi.org/10.1371/journal.pone.0082754}{{\em PloS one} {\bf 8}
  (Jan., 2013)  e82754}, \href{http://arxiv.org/abs/1303.6596}{{\tt
  arXiv:1303.6596 [cond-mat.stat-mech]}}.

\bibitem{Zulkowski2014bit}
P.~R. Zulkowski and M.~R. DeWeese, ``{Optimal finite-time erasure of a
  classical bit},'' \href{http://dx.doi.org/10.1103/PhysRevE.89.052140}{{\em
  Phys. Rev. E} {\bf 89} (May, 2014)  052140},
  \href{http://arxiv.org/abs/1310.4167}{{\tt arXiv:1310.4167
  [cond-mat.stat-mech]}}.

\bibitem{Mandal2015nessmet}
D.~Mandal and C.~Jarzynski, ``{Analysis of slow transitions between
  nonequilibrium steady states},'' \href{http://arxiv.org/abs/1507.06269}{{\tt
  arXiv:1507.06269 [cond-mat.stat-mech]}}.

\bibitem{amari2007methods}
S.-i. Amari and H.~Nagaoka, {\em Methods of information geometry}, vol.~191.
\newblock American Mathematical Soc., 2007.

\bibitem{Cramer1945}
H.~Cram{\'{e}}r, {\em {Mathematical Methods of Statistics}}.
\newblock Princeton University Press, 1945.

\bibitem{RadhakrishnaRao1945}
C.~{Radhakrishna Rao}, ``{Information and the accuracy attainable in the
  estimation of statistical parameters},'' {\em Bull. Calcutta Math. Soc.} {\bf
  37} (1945)  81--91.

\bibitem{Glauber1963}
R.~J. Glauber, ``{Time-Dependent Statistics of the Ising Model},''
  \href{http://dx.doi.org/10.1063/1.1703954}{{\em Journal of Mathematical
  Physics} {\bf 4} (Dec., 1963)  294}.

\bibitem{Skoge2011coop}
M.~Skoge, Y.~Meir, and N.~S. Wingreen, ``{Dynamics of cooperativity in chemical
  sensing among cell-surface receptors},''
  \href{http://dx.doi.org/10.1103/PhysRevLett.107.178101}{{\em Phys. Rev.
  Lett.} {\bf 107} (oct, 2011)  1--5},
  \href{http://arxiv.org/abs/1109.4160}{{\tt arXiv:1109.4160 [q-bio.MN]}}.

\bibitem{Hohenberg1977critdyn}
P.~C. Hohenberg and B.~I. Halperin, ``{Theory of dynamic critical phenomena},''
  \href{http://dx.doi.org/10.1103/RevModPhys.49.435}{{\em Reviews of Modern
  Physics} {\bf 49} (jul, 1977)  435--479}.

\bibitem{Ganguli2008}
S.~Ganguli, D.~Huh, and H.~Sompolinsky, ``{Memory traces in dynamical
  systems.},'' \href{http://dx.doi.org/10.1073/pnas.0804451105}{{\em Proc.
  Natl. Acad. Sci. U.S.A.} {\bf 105} (dec, 2008)  18970--5}.

\bibitem{Ganguli2010}
S.~Ganguli and H.~Sompolinsky, ``{Short-term memory in neuronal networks
  through dynamical compressed sensing},'' in {\em Advances in Neural
  Information Processing Systems}, pp.~667--675.
\newblock 2010.

\bibitem{kemeny1960finite}
J.~Kemeny and J.~Snell, {\em Finite markov chains}.
\newblock Springer, 1960.

\bibitem{Hatano2001}
T.~Hatano and S.-i. Sasa, ``{Steady-State Thermodynamics of Langevin
  Systems},'' \href{http://dx.doi.org/10.1103/PhysRevLett.86.3463}{{\em Phys.
  Rev. Lett.} {\bf 86} (Apr., 2001)  3463--3466},
  \href{http://arxiv.org/abs/cond-mat/0010405}{{\tt arXiv:cond-mat/0010405
  [cond-mat.stat-mech]}}.

\bibitem{Lawler1988cheeger}
G.~F. Lawler and A.~D. Sokal, ``Bounds on the $L^2$ Spectrum for Markov Chains
  and Markov Processes: A Generalization of Cheeger's Inequality,''
  \href{http://dx.doi.org/10.2307/2000925}{{\em Transactions of the American
  Mathematical Society} {\bf 309} (1988) no.~2, 557--580}.

\bibitem{Mayer2004}
P.~Mayer and P.~Sollich, ``{General solutions for multispin two-time
  correlation and response functions in the Glauber–-Ising chain},''
  \href{http://dx.doi.org/10.1088/0305-4470/37/1/002}{{\em Journal of Physics
  A: Mathematical and General} {\bf 37} (Jan., 2004)  9--49},
  \href{http://arxiv.org/abs/cond-mat/0307214}{{\tt arXiv:cond-mat/0307214
  [cond-mat.stat-mech]}}.

\bibitem{Skoge2013}
M.~Skoge, S.~Naqvi, Y.~Meir, and N.~S. Wingreen, ``{Chemical Sensing by
  Nonequilibrium Cooperative Receptors.},''
  \href{http://dx.doi.org/10.1103/PhysRevLett.110.248102}{{\em Phys. Rev.
  Lett.} {\bf 110} (June, 2013)  }, \href{http://arxiv.org/abs/1307.2930}{{\tt
  arXiv:1307.2930 [q-bio.MN]}}.

\end{thebibliography}\endgroup
\bibliographystyle{utcaps_sl}

\end{document}